\documentclass[english,aps,pre,twocolumn,superscriptaddress,showpacs]{revtex4-1}
\usepackage[T1]{fontenc}
\usepackage{array}
\usepackage{verbatim}
\usepackage{amsmath}
\usepackage{amssymb}
\usepackage{graphicx}
\usepackage{esint}
\usepackage{epstopdf}
\usepackage{hyperref}
\usepackage{color}

\makeatletter

\providecommand{\tabularnewline}{\\}

\@ifundefined{textcolor}{}
{%
 \definecolor{BLACK}{gray}{0}
 \definecolor{WHITE}{gray}{1}
 \definecolor{RED}{rgb}{1,0,0}
 \definecolor{GREEN}{rgb}{0,1,0}
 \definecolor{BLUE}{rgb}{0,0,1}
 \definecolor{CYAN}{cmyk}{1,0,0,0}
 \definecolor{MAGENTA}{cmyk}{0,1,0,0}
 \definecolor{YELLOW}{cmyk}{0,0,1,0}
}

\makeatother

\usepackage{babel}
\begin{document}

\preprint{Version 1, \today}

\title{Control of diffusion of nano-particles in an optical vortex lattice}

\author{Ivar Zapata}

\email{izapatao@ucm.es}

\selectlanguage{english}%

\affiliation{Condensed Matter Physics Center (IFIMAC),  Universidad Aut\'{o}noma de Madrid, 28049, Madrid, Spain.}

\affiliation{Departamento Física de Materiales,
 Universidad Complutense de Madrid,  Madrid, Spain}

\author{Rafael Delgado-Buscalioni}


\selectlanguage{english}%


\affiliation{Condensed Matter Physics Center (IFIMAC),  Universidad Aut\'{o}noma de Madrid, 28049, Madrid, Spain.}

\author{Juan José Sáenz}

\affiliation{Donostia International Physics Center (DIPC), Paseo Manuel Lardizabal 4, 20018 Donostia-San Sebastian, Spain.}

\affiliation{IKERBASQUE, Basque Foundation for Science, 48013 Bilbao, Spain.}

 
%


\selectlanguage{english}%


\begin{abstract}
A two-dimensional periodic optical force field, which combines conservative dipolar forces with vortices from radiation pressure, is proposed in order to influence the diffusion properties of optically susceptible nano-particles. The different deterministic flow patterns are identified. In the low noise limit, the diffusion coefficient is computed from a Mean First Passage Time (MFPT) and the Most Probable Escape Paths (MPEP) are identified for those flow patterns which possess an stable stationary point. Numerical simulations of the associated Langevin equations show remarkable agreement with the analytically deduced expressions. Modifications of the force field are proposed so that a wider range of phenomena could be tested.
\end{abstract}
\maketitle

\section{Introduction}\label{sec:introduction}

The study of  Brownian motion in spatially periodic and random  landscapes has long been a topic of interest in all branches of science \cite{Risken1984,bouchaud1990anomalous,sahimi1993flow,reimann2002brownian,hanggi2009artificial,evers2013colloids}. 
Counterintuitive phenomena like the giant diffusion induced by an oscillating periodic potential \cite{reimann2001giant} or kinetically locked-in states in driven diffusive transport \cite{Reichhardt99p414} have been realized by using optically induced potential energy  landscapes \cite{lee2006giant,Korda02p128301} demonstrating optical guiding and sorting of particles in microfluidic flows  \cite{MacDonald05p421,xiao2010sorting}.
Fascinating complex dynamics appear when thermal Brownian motion is combined with additional energy sources \cite{reimann2002brownian,hanggi2009artificial,erdmann2000brownian,howse2007self}. 

Active Brownian objects are able of taking up energy from their environment and converting it into directed motion that drives them out of equilibrium. The energy necessary for active motion can  be supplied by an  external spatio-temporal  modulated field or by energy input from a local ``self-generated'' force  \cite{hanggi2009artificial} (
small objects can swim by generating  concentration and other gradients around them   \cite{paxton2005motility,golestanian2007designing}). 
Active Brownian motion can be encountered in a large variety of phenomena, including protein diffusion 
\cite{gambin2006lateral},   the motion of swimming bacteria \cite{berg2008coli,baskaran2009statistical}
and artificial nanoscale swimming devices \cite{golestanian2007designing,ebbens2010pursuit},  turbulent flows \cite{shraiman2000scalar,eyink2006onsager} or even in the processes leading to collective opinion formation   \cite{Schweitzer00p723}.  
Self-generated forces can also be optically tuned  through photo-chemical \cite{yashin2012mechano} or photo-thermal \cite{golestanian2012collective,buttinoni2012active} mechanisms. Direct conversion of electromagnetic energy into mechanical energy can be achieved by time modulation of periodic or random intensity patterns. Optically induced active superdiffusion can be realized in  colloidal suspensions where the  particles move in a time-dependent, self-generated, speckle intensity pattern  induced by multiple scattering of light \cite{douglass2012superdiffusion}. Active superdiffusion can also be achieved by time-varying external speckle patterns changing over timescales similar to the characteristic times of thermal Brownian fluctuations \cite{volpe2014brownian} (slower, adiabatic, changes leading to subdiffusive behavior \cite{volpe2014brownian} while faster changes can lead to conservative, Casimir-like, interactions between particles \cite{bruegger2015controlling}).

Most of previous work on thermal ratchets and control of active media relied on potential energy landscapes with
some type of external, random or periodic, time-dependent driving.  However, there is a way of driving the system out of equilibrium which does not require time-dependent forces, namely, the use of non-conservative stationary force fields with a nonvanishing solenoidal component, i.e. a component which can be expressed as the curl of a vector \cite{roichman2008influence,sun2009brownian,albaladejo2009scattering}. Even in absence of thermal noise, the curl dynamics of systems under Newtonian curl forces present intriguing properties that are not yet fully understood \cite{berry2013physical,berry2015hamiltonian} (note that the term ``curl force'' is also being used to denote a force field whose curl is different from zero). 

The practical realization of Brownian vortexes \cite{roichman2008influence,sun2009brownian}  driven by curl forces on particles trapped in  optical tweezers  \cite{Ashkin97p4853,Jones2015} 
has generated an increasing interest on the study of nonequilibrium dynamics of optically trapped particles 
\cite{pesce2009quantitative,wu2009direct,iglesias2011scattering,divitt2015cancellation}. 
Interesting non-equilibrium effects, like giant diffusion \cite{Albaladejo2009,albaladejo2011light} or deterministic  ratchet effects \cite{Zapata2009deterministic}, have been predicted for the active Brownian diffusion in two-dimensional periodic  curl force fields.    However, theoretical description of the diffuse motion of objects  in stationary periodic curl force landscapes remains to be done. 

Since the early works on reaction rates by Kramer (for a review see \cite{Hanggi1990}) most of the multi-dimensional studies have dealt with situations in which the deterministic driving forces derive from a potential scalar (but see for example \cite{Filliger2007} for a case in which inertia together with a magnetic force is studied). 
When the forces are stationary and non-conservative, the construction of the equilibrium distribution poses a formidable problem. Early works in this direction (see the review in \cite{Graham1987a}) pointed that the essential difficulty is of geometric nature. The characteristics of this stationary distribution are needed in order to asymptotically estimate the transitions rates in the regime of low noise, which is, computationally, the most elusive. In this regime, it is not difficult to show that when the underlying force field is periodic in space, diffusion can be formulated as a first passage time (FPT) problem, therefore bringing reaction rates into discussion when a diffusion coefficient is needed.

FPT problems can be most simply described as finding the distribution of times according to which a random process first exceeds a prescribed threshold or reaches a specified configuration. Although 
we lack a general theory to study these problems in more than two dimensions, the 2D case seems to have made consistent progress in the last two decades. The low temperature FPT is mostly determined by solving an associated 4D Hamiltonian (two space and two momenta coordinates), whose zero energy solutions can be organized into 2D Lagrangian surfaces. The geometric understanding of these surfaces is the most important part of the study of low noise FPT. In \cite{Maier1996} it is shown that focusing and caustics occur in Lagrangian surfaces and their emergence modifies the naive FPT Arrhenius estimates coming from Eyring formula. When there is no focusing, a new classically forbidden wedge region at one side of the exit saddle points is discovered in \cite{Maier1997}, which also puts Eyring formula on a firm ground for the first time. For a variety of related new results, see \cite{Dykman1994a,Dykman1994,Smelyanskiy1997,Smelyanskiy1997a,Dykman1999}. Non-mathematical applications of these theoretical results can be found in \cite{Luchinsky1997,Luchinsky1997a,Guerin2011,Bressloff2013}. They include analogue electrical circuits or highly stylized biological or neural network systems.  

Our purpose here is twofold: on a first hand,
we propose to apply the FPT results from \cite{Maier1996,Maier1997} in order to compute the diffusion characteristics of over-damped nano-particles on a  two-dimensional periodic  curl force field of an optical lattice 
(see Ref.  \cite{Siler2010} for an example of application of the FPT approach to describe the diffusion of nanoparticles on -conservative- one-dimensional optical lattices).  On the other hand, 
since this is a relatively  simple and  experimentally feasible physical system, we will show that  simple modifications of the proposed experimental setup could allow to physically realise and explore most of  the full spectrum of novel results, i.e., focusing, caustics and classically forbidden wedge regions. The present setups are experimentally highly controllable and the mathematical models which describe their physics can perform quantitative and precise predictions. 

Our paper is organized as follows. In Sec. \ref{sec:physical-model} we review the model and introduce Langevin equations. Sec. \ref{sec:features-of-the-deterministic-flow} analyzes the features of the deterministic flow and shows a complete classification of the topologically distinct flow patterns. In Sec. \ref{sec:fokker-planck-equation-description} we review the Fokker-Planck approach to diffusion and its relation the Mean First Passage Time (MFPT). Sec. \ref{sub:Low-noise-expansion.}, which deals with the low-noise approximation to the MFPT, is a summary of the results from \cite{Maier1997} needed to understand the present work. The application of these results to the model discussed in the present work is given in Sec. \ref{sec:analytical-results-for-the-physical-system.-lack-of-focusing.}. Sec. \ref{sub:Conservative-systems} summarizes relevant formulas for the diffusion coefficient in limit cases in which the non-conservative forces do not appear, and Sec. \ref{sec:no-stable-nodes} provides the discussion on systems without any stable node. Numerical simulation of Langevin equations and its successful comparison with  low-noise expansion results is presented in Sec. \ref{sec:comparison-with-numerical-langevin-simulations.}. Finally, we conclude in Sec. \ref{sec:conclusions}, with an outlook of possible modifications of the setup in order to probe those other low-noise asymptotics phenomena mentioned in the introduction and discussed in Sec. \ref{sub:Low-noise-expansion.}.

\section{Physical model}\label{sec:physical-model}

The model is described in \citet{Albaladejo2009}. Spherical nano-particles
are confined in a plane with coordinates $\mathbf{R}=\left(X,Y\right)$
and are subject to light forces coming from the interference of two
standing waves propagating along the directions $x$ and $y$ respectively,
polarized along a direction, $Z$, orthogonal to the plane of motion
of the nano-particles. The two standing waves have the same frequency
$\omega$ and intensity. The $Z$ component of the electric field
amplitude in the interference region can be written as
\begin{equation}
E_{z}=i2E_{0}\left(\sin\left(kX\right)+e^{i\phi}\sin\left(kY\right)\right),\label{eq:ElectricField}
\end{equation}
where $k=n\omega/c$ is the wave-number, $c=\sqrt{1/\varepsilon_{0}\mu_{0}}$
is the speed of light in vacuum, $n=\sqrt{\varepsilon}$ the refraction
index of the surrounding medium with $\varepsilon$ as relative electric
permittivity, $E_{0}$ is the common amplitude of the electric field
of the two standing waves, related to the laser power density $P$
through $\left(n/c\right)P=\varepsilon_{0}\varepsilon|E_{0}|^{2}/2$,
and $\phi$ is the phase different between both standing waves. Then,
nano-particles are subject to a light force $\mathbf{F}$ which can
be decomposed into a conservative part $\mathbf{F}_{cons}$ and a
non-conservative $\mathbf{F}_{n-cons}$ coming from radiation pressure:
\begin{eqnarray}
\mathbf{F} & = & \mathbf{F}_{cons}+\mathbf{F}_{n-cons}\nonumber \\
\mathbf{F}_{cons} & = & -\mathbf{\nabla}U\left(X,Y\right)\nonumber \\
\mathbf{F}_{n-cons} & = & \mathbf{\nabla\times}\left[\mathbf{z}A\left(X,Y\right)\right]\label{eq:Force}
\end{eqnarray}
with $\mathbf{z}$ a unit vector in the $z$ direction and $U\left(x,y\right)$
and $A\left(x,y\right)$ are given by
\begin{eqnarray}
U\left(X,Y\right) & = & -2\frac{n}{c}P\alpha'\left[\sin^{2}\left(kX\right)+\sin^{2}\left(kY\right)\right. \nonumber \\
 &  & + \left. 2\cos\left(\phi\right)\sin\left(kX\right)\sin\left(kY\right)\right] \nonumber  \\
A\left(X,Y\right) & = & 4\frac{n}{c}P\alpha''\sin\left(\phi\right)\cos\left(kX\right)\cos\left(kY\right)\label{eq:Potentials}
\end{eqnarray}
where $\alpha\left(\omega\right)=\alpha'+i\alpha''$ is the complex
polarizability of the nano-particles. Within the electric dipole approximation,
$a\ll2\pi/k$, ($a$ is the radius of spherical nano-particles), the
polarizability is given by $\alpha\left(\omega\right)=\alpha_{0}\left(\omega\right)\left(1-i\alpha_{0}\left(\omega\right)k^{3}/6\pi\right)^{-1}$,
$\alpha_{0}\left(\omega\right)=4\pi a^{3}\left(\varepsilon_{np}\left(\omega\right)-\varepsilon\right)\left(\varepsilon_{np}\left(\omega\right)+2\varepsilon\right)^{-1}$
is the usual instantaneous dielectric response of dielectric spheres
(Clausius-Mossotti relation) and $\varepsilon_{np}\left(\omega\right)$
is the macroscopic relative permittivity of the substance of which
nano-particles are made off. The correction $1\rightarrow\left(1-i\alpha_{0}\left(\omega\right)k^{3}/6\pi\right)^{-1}$
to the polarizability comes from radiation reaction and is sufficient
to enforce the validity of the optical theorem in scattering when
the electric dipole approximation is made (see \cite{albaladejo2010radiative,Novotny2012}).

In this work a surrounding medium of viscosity $\eta$ is considered.
Nano-particles' friction coefficient, $\gamma$, which appears in
Newton's force equation as $-\gamma d\mathbf{R}/dT$, is given by
Stokes' law $\gamma=6\pi a\eta$. The inertia term $md^{2}\mathbf{R}/dT^{2}$
can be neglected in the so called over-damped regime, which here means
that the scale of the optical forces is sufficiently small $k\sqrt{8m\alpha'P\left(n/c\right)}\ll\gamma$.
The effective over-damped Langevin equation of motion for the independent
nano-particles can be written as
\begin{eqnarray}
\gamma\frac{d\mathbf{R}}{dT} & = & \mathbf{F}+\mathbf{\Xi}\left(T\right)\nonumber \\
\left\langle \Xi_{i}\left(T_{1}\right)\Xi_{j}\left(T_{2}\right)\right\rangle  & = & 2\gamma k_{B}T\delta_{ij}\delta\left(T_{1}-T_{2}\right),\label{eq:LangevinEquationDimensional}
\end{eqnarray}

Using the following change of variables and definitions,
\begin{eqnarray}
\mathbf{r} & = & k\mathbf{R}\nonumber \\
t & = & T/\tau\nonumber \\
\tau & := & \frac{\gamma}{2\left(n/c\right)P|\alpha'|k^{2}}\nonumber \\
\xi\left(t\right) & := & \frac{\tau}{\gamma}\Xi\left(T\right)\nonumber \\
\beta & := & \frac{\alpha''}{|\alpha'|}\ge0
\end{eqnarray}
the Langevin equation (Eq. \ref{eq:LangevinEquationDimensional}), can be
written in dimensionless form
\begin{equation}
\frac{d\mathbf{r}}{dt}=\mathbf{f}+\mathbf{\xi}\left(t\right)\label{eq:LangevinEquation}
\end{equation}
where the non-dimensional optical force becomes,
\begin{eqnarray}
\mathbf{f} &=& \pm\mathbf{\nabla}\left(\sin^{2}x+\sin^{2}y+2\cos\phi\sin x\sin y\right)\nonumber \\
& &+2\beta\sin\phi\mathbf{\nabla\times}\left(\mathbf{z}\cos x\cos y\right) 
\end{eqnarray}
and the noise and noise amplitude parameter $\epsilon$ are given
by,
\begin{eqnarray}
\left\langle \xi_{i}\left(t_{1}\right)\xi{}_{j}\left(t{}_{2}\right)\right\rangle  & = & \epsilon\delta_{ij}\delta\left(t{}_{1}-t{}_{2}\right)\nonumber \\
\epsilon & := & \frac{k_{B}T}{\left(n/c\right)P|\alpha'|}.\label{eq:noiseAmp}
\end{eqnarray}
Here the gradient $\mathbf{\nabla}$ is with respect to $\mathbf{r}$
and the $\pm$ is the sign of $\alpha'$. In the present work we will
concentrate on the normal $\alpha'>0$, the opposite case will be
done in a further research work. The drift force $\mathbf{f}$ can
be also written in matrix form,
\begin{eqnarray}
\mathbf{f} & = & 2\left[\begin{array}{c}
\cos x\left(\sin x+c_{-}\sin y\right)\\
\cos y\left(\sin y+c_{+}\sin x\right)
\end{array}\right]\label{eq:drift}\\
c_{\pm} & := & \cos\phi\pm\beta\sin\phi.\nonumber
\end{eqnarray}
We will see that we can take $0\le\phi\le\pi/2$ without loosing any
generality. This implies the following constrains:
\begin{eqnarray*}
0 & \le & c_{-}+c_{+}\le2\\
c_{-} & \le & c_{+}.
\end{eqnarray*}
It is not difficult to show that once these two constraints are fulfilled,
then both $\phi,\beta$ exist ($0\le\phi\le\pi/2,\,\beta\ge0$) so that $c_{\pm}=\cos\phi\pm\beta\sin\phi$.
To see that it is sufficient to have $0\le\phi\le\pi/2$, note that both $\left(\cos\phi,\sin\phi\right)\Rightarrow\left(\cos\phi,-\sin\phi\right),\,\left(x,y\right)\Rightarrow\left(y,x\right)$
and $\left(\cos\phi,\sin\phi\right)\Rightarrow\left(-\cos\phi,\sin\phi\right),\,\left(x,y\right)\Rightarrow\left(-y,x\right)$
transform the equations of motion into one in which $0\le\phi\le\pi/2$.
It is finally noted that $\epsilon$, the ratio between thermal
energy and work done by the laser on the particle, is in fact
the inverse of Peclet number controlling the advection-diffusion
dynamics.

\section{Features of the deterministic flow}\label{sec:features-of-the-deterministic-flow}

\begin{table*}
	\begin{tabular}{|p{0.8cm}|p{9cm}|}
		\hline
		\texttt{s1}  & $\mathbf{r}_{s1}=\left(2n+1,2m+1\right)\pi/2$\tabularnewline
		\hline
		\texttt{s2}  & $\mathbf{r}_{s2}=\left(n,m\right)\pi$\tabularnewline
		\hline
		\texttt{s3}  & $\mathbf{r}_{s3}=\left(\left(2n+1\right)\pi/2,\left(-1\right)^{n+1}\arcsin c_{+}\right)$\tabularnewline
		\hline
		\texttt{s4}  & $\mathbf{r}_{s4}=\left(\left(-1\right)^{m+1}\arcsin c_{-},\left(2m+1\right)\pi/2\right)$\tabularnewline
		\hline
		& $\phi=0$ and both lines $x+y=2n\pi$ and $x-y=(2m+1)\pi$\tabularnewline
		\hline
	\end{tabular}
	\caption{\label{tab:Stationary-points-of} Stationary points of deterministic
		flow $\mathbf{\dot{r}=f\left(r\right)}$ and their notation (see Eq. \ref{eq:drift})}
\end{table*}

The main objective of this work is to analyze the most probable route(s)
of the particle over the periodic landscape of forces exerted by the
lasers. To that end we first analyze the symmetries of the force pattern
and the types of stationary points (where the driving force vanishes
${\bf f}=0$). It is noted that the additional presence of a minute
amount of thermal noise is essential to let the particle scape from
the \emph{stable} stationary points, which in this case can be termed
as \emph{metastable}. This low noise limit is the subject of the present
work, which focus on how long it takes the particle to scape and where (as discussed in Sec. \ref{sec:Diffusion}), so as to provide control over the long time particle diffusion.

\subsection{Symmetries\label{sub:Symmetries}}

Obviously, the flow is translational invariant in units of $2\pi$,
both vertically or horizontally, therefore we can restrict to a square
$\left(-\pi,\pi\right)^{2}$. A special status is given to the lines
$x=\left(2n+1\right)\pi/2$ and $y=\left(2m+1\right)\pi/2$, where
the force is respectively vertical or horizontal (here and in the
rest of the present work, $n,m\in\mathbb{Z}$ and note that $m$ should
not be confused with the mass). In fact, it can be shown that horizontal
and vertical reflections through any of these lines together with
inversion through their crossings, i.e., at the points $\left(2n+1,2m+1\right)\pi/2$,
are symmetries (the equations of motion are invariant). Inversion
through $\left(n,m\right)\pi$ (in particular through the origin)
are symmetries as well.

\subsection{Classification of stationary points\label{sub:Classification-of-stationary}}

The full set of stationary points $\mathbf{f}\left(\mathbf{r}_{s}\right)=\mathbf{0}$,
of the drift are shown in Table \ref{tab:Stationary-points-of} and
also illustrated in Fig.\ref{fig:grDrifTypeE}-\ref{fig:grDriftCodesST}.

Obviously, for the fixed points of type \texttt{s3} and \texttt{s4}
($\mathbf{r}_{s3},\mathbf{r}_{s4}$) it is required that either $|c_{+}|\le1$
or $|c_{-}|\le1$, and the $\arcsin$ functions then refer to all
their possible values, not just their principal branch definitions.

The character of the flows near these fixed points can be obtained
from the linearized drift $\mathbf{f}_{i}\left(\mathbf{r}-\mathbf{r}_{s}\right)\simeq\mathbf{B}\left(\mathbf{r}_{s}\right)_{ij}\left(\mathbf{r}-\mathbf{r}_{s}\right)_{j}$,
with $\mathbf{B}\left(\mathbf{r}_{s}\right)_{ij}:=\mathbf{\nabla}_{j}\mathbf{f}_{i}\left(\mathbf{r}_{s}\right)$
(here we use Einstein summation convention on repeated indices and
$i,j=1,2$ with $\mathbf{r}_{1}=x$, etc). The fixed points nomenclature
is explained in \citet{Almeida1990}. Here we find the following possibilities
for our 2D system, depending on the eigenvalues of $\lambda_{1,2}$
of $\mathbf{B}\left(\mathbf{r}_{s}\right)$: stable (SN) $\lambda_{1,2}<0$
or unstable (UN) $\lambda_{1,2}>0$ node, saddle (SD) $\lambda_{1}<0<\lambda_{2}$
and unstable focus (UF) $\Re\left(\lambda_{1,2}\right)>0,\,\Im\left(\lambda_{1,2}\right)\ne0$.
Table \ref{tab:Flow-types} summarizes the possible qualitatively
different flows.

\begin{table*}
\begin{tabular}{|c|c|c|c|c|c|}
\hline
Case  & Range  & $\mathbf{r}_{s1}$  & $\mathbf{r}_{s2}$  & $\mathbf{r}_{s3}$  & $\mathbf{r}_{s4}$\tabularnewline
\hline
\hline
A  & $0\le c_{-}<c_{+}<1$  & SN  & UN  & SD  & SD\tabularnewline
\hline
B  & $-1<c_{-}<0<c_{+}<1$  & SN  & UF  & SD  & SD\tabularnewline
\hline
Impossible  & $c_{-}<-1,\,0<c_{+}<1$  &  &  &  & \tabularnewline
\hline
C  & $0\le c_{-}<1<c_{+}$  & SN/SD({*})  & UN  &  & SD\tabularnewline
\hline
D  & $-1<c_{-}<0\,,1<c_{+}$  & SN/SD({*})  & UF  &  & SD\tabularnewline
\hline
E  & $c_{-}<-1,\,1<c_{+}$  & SD  & UF  &  & \tabularnewline
\hline
\multicolumn{6}{|c|}{({*})$n+m$ even/odd }\tabularnewline
\hline
\end{tabular}
\caption{\label{tab:Flow-types}Flow types. The available character of the
	fixed points are: SN stable node, UN unstable node, SD saddle and
	UF unstable focus. }
\end{table*}

The cases in which $c_{+}=c_{-}$ lead to vanishing non-conservative
forces: we will deal with them separately in Sec. \ref{sub:Conservative-systems}.
On the other hand, in the present work we will not consider bifurcation
phenomena which arises when $|c_{-}|=1$ or $|c_{+}|=1$ so that 
stable nodes change into saddles, i.e., one of the eigenvalues $\lambda_{1,2}$ 
goes through zero. In these situations, linear analysis is not sufficient to decide
the stability of the corresponding stationary points.

In the subsequent analysis, a special role will be given to both stable
nodes (SN) and the saddles (SD), the later will be the only possible
candidates through which particle flow will pass in the low temperature
limit ($\epsilon\ll1$). For the saddles it is necessary to obtain
both stable and unstable eigenvalues and eigenvectors, $\lambda_{s/u}$
and $\mathbf{e}_{s/u}$ (we will explain later the meaning of the
symbols $\left(\tilde{\mathbf{e}}_{s},\tilde{\mathbf{g}}_{s}\right)$
and $\left(\tilde{\mathbf{e}}_{u},\tilde{\mathbf{g}}_{u}\right)$).
Table \ref{tab:Unstable/stable--eigenvalues} provides the eigenvectors
and eigenvalues of the cases of Table \ref{tab:Flow-types} where
saddles can occur.

\begin{table*}
\begin{tabular}{|c|c|c|c|c|c|c|}
\hline
SD  & $\lambda_{s}$  & $\lambda_{u}$  & $\mathbf{e}_{s}$  & $\mathbf{e}_{u}$  & $\left(\tilde{\mathbf{e}}_{s},\tilde{\mathbf{g}}_{s}\right)$  & $\left(\tilde{\mathbf{e}}_{u},\tilde{\mathbf{g}}_{u}\right)$\tabularnewline
\hline
\hline
$\mathbf{r}_{s1}$, $n+m$ odd & $-2\left(1-c_{-}\right)$  & $2\left(c_{+}-1\right)$  & $\left(1,0\right)$  & $\left(0,1\right)$  & $\left(0,-1/2\lambda_{u},0,1\right)$  & $\left(-1/2\lambda_{s},0,1,0\right)$\tabularnewline
\hline
$\mathbf{r}_{s1}$, $n+m$ even  & $-2\left(1+c_{+}\right)$  & $-2\left(1+c_{-}\right)$  & $\left(0,1\right)$  & $\left(1,0\right)$  & N. A.   & N. A.\tabularnewline
\hline
$\mathbf{r}_{s3}$  & $-2\left(1-c_{-}c_{+}\right)$  & $2\left(1-c_{+}^{2}\right)$  & $\left(1,0\right)$  & $\left(0,1\right)$  & $\left(0,-1/2\lambda_{u},0,1\right)$  & $\left(-1/2\lambda_{s},0,1,0\right)$\tabularnewline
\hline
$\mathbf{r}_{s4}$  & $-2\left(1-c_{-}c_{+}\right)$  & $2\left(1-c_{-}^{2}\right)$  & $\left(0,1\right)$  & $\left(1,0\right)$  & $\left(-1/2\lambda_{u},0,1,0\right)$  & $\left(0,-1/2\lambda_{s}0,1\right)$\tabularnewline
\hline
\end{tabular}
\caption{\label{tab:Unstable/stable--eigenvalues}Unstable/stable $\lambda_{u/s}$
	eigenvalues and eigenvectors $\mathbf{e}_{s/u}$ of the saddles together
	with they eigenvectors $\left(\tilde{\mathbf{e}}_{s},\tilde{\mathbf{g}}_{s}\right)$
	and $\left(\tilde{\mathbf{e}}_{u},\tilde{\mathbf{g}}_{u}\right)$
	corresponding to stable and unstable direction in the 4D phase space
	of the effective Hamiltonian dynamics (see Eq. \ref{eq:linearizedHamilton}).}
\end{table*}

\subsection{Qualitative flow regimes}\label{sec:qualitative-flow-regimes}

\begin{figure}
\includegraphics[width=1\columnwidth]{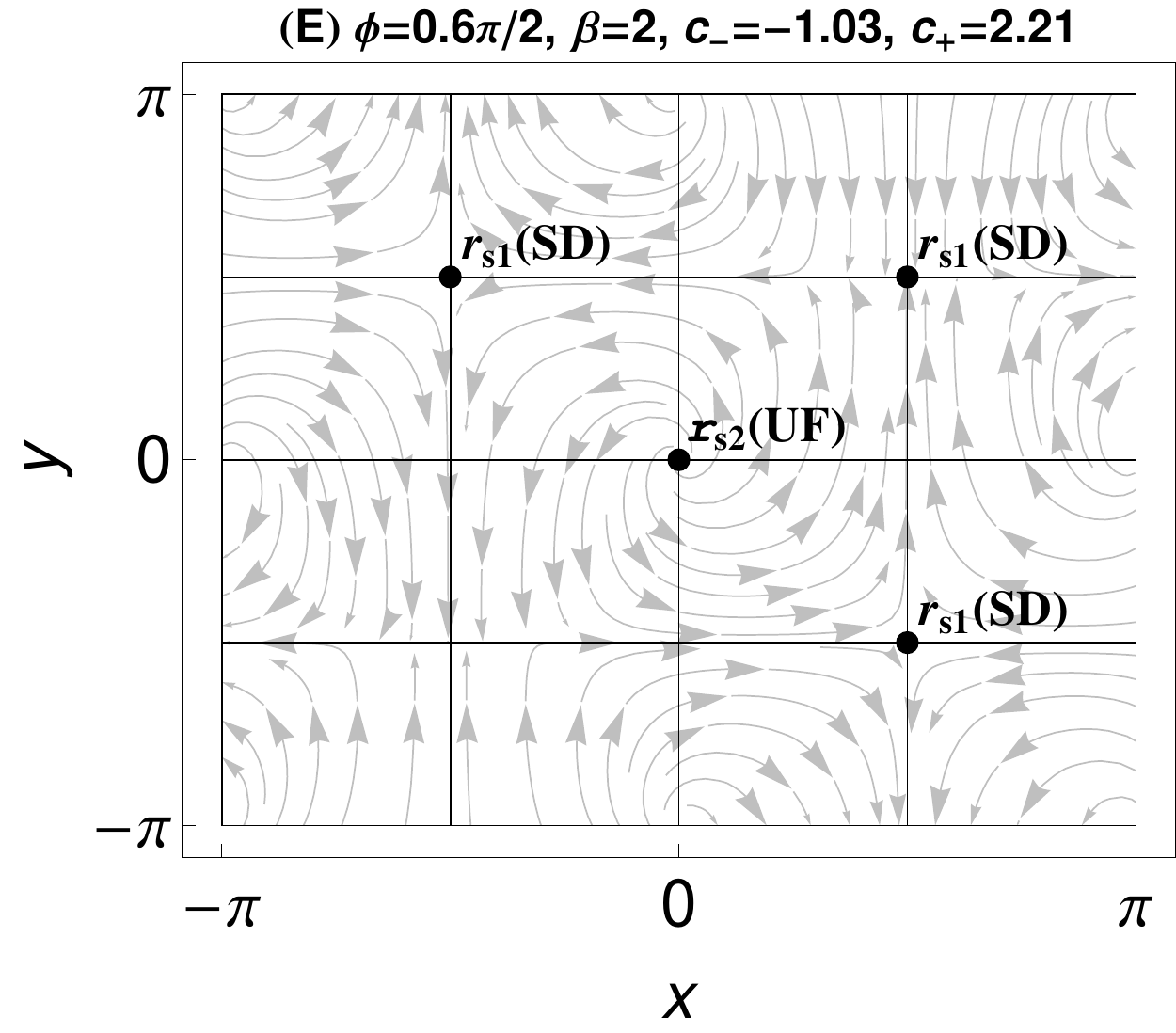}

\caption{\label{fig:grDrifTypeE}An example of flow of type E). Symmetry transformations
(see \ref{sub:Symmetries}) applied to the marked stationary points
leads to all the stationary points. }
\end{figure}

\begin{figure*}
\includegraphics[width=2\columnwidth]{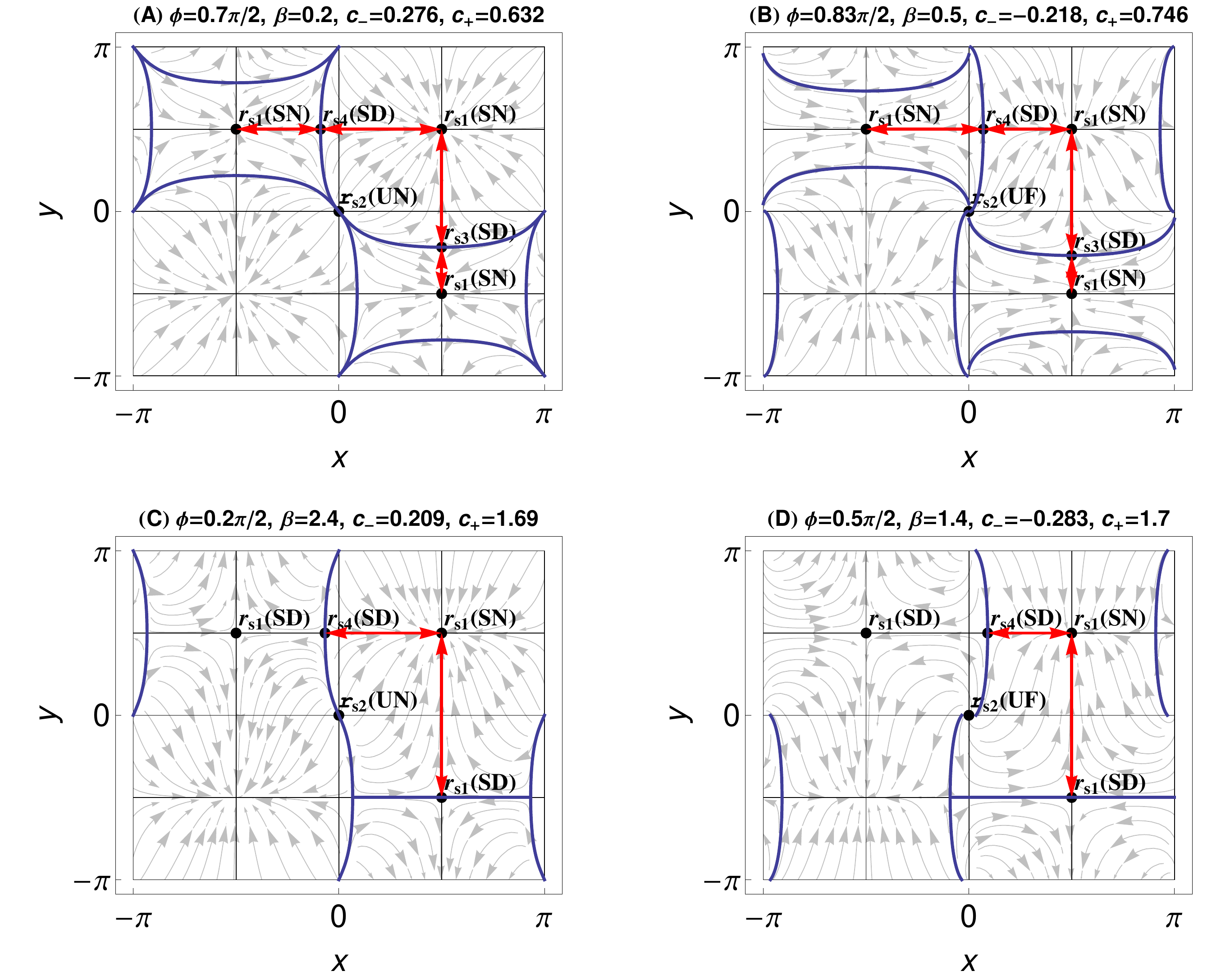}

\caption{\label{fig:grDriftCodesST} Examples of flows types A) through D)
(see Table \ref{tab:Flow-types}). The deterministic flow lines from
$\mathbf{\dot{r}=f\left(r\right)}$ (see Eq.\ref{eq:drift}) are shown
in gray, the stationary points are black dots, labeled by their character
(see Table \ref{tab:Stationary-points-of}). Symmetry transformations
(see \ref{sub:Symmetries}) applied to the marked stationary points
leads to all the stationary points.The separatrices, delimiting the
regions of attraction from the stable nodes, are shown in blue. The
red arrowed lines denote the MPEP candidates.}
\end{figure*}

From the results in \ref{sub:Classification-of-stationary} and Table
\ref{tab:Flow-types} it can be seen that cases A) through D) lead
to stable nodes. This contrasts with case E) in which there are no
attractive stationary points, and where the limit sets of the spiral surrounding
unstable foci at $\mathbf{r}_{s2}$ are heteroclinic orbits joining
the SD's at the four corners, located at points of type \texttt{s1}
($\mathbf{r}_{s1}$). An example of case E) is shown in Fig.\ref{fig:grDrifTypeE}.
When $\phi=\pi/2$ and $\beta>1$, due to the small distance among
neighbor limit cycles, diffusion is gigantically enhanced, as is shown
in \citet{Albaladejo2009}. We will consider case E) in Sec. \ref{sec:no-stable-nodes}, but most of the results will be about regimes with
stable nodes: the corresponding flows are illustrated in Fig. \ref{fig:grDriftCodesST}.

\section{Diffusion in the vortex field\label{sec:Diffusion}}

\subsection{Fokker-Planck equation description}\label{sec:fokker-planck-equation-description}

In a periodic 2D static force field, like the one described in the
Langevin equation (Eq. \ref{eq:LangevinEquation}), an effective diffusion
coefficient can be defined as
\begin{equation}
D:=\lim_{t\rightarrow\infty}\left\langle \frac{|\mathbf{r}\left(t\right)-\mathbf{r}\left(0\right)|^{2}}{4t}\right\rangle ,\label{eq:Diffusion}
\end{equation}
where the average is taken over those realizations starting with the same
initial position $\mathbf{r}\left(0\right)$. This formula requires,
that no average drift is present, $\lim_{t\rightarrow\infty}\left\langle \mathbf{r}\left(t\right)\right\rangle /t=\mathbf{0}$,
a requirement which is fulfilled by the present forces, Eq. \ref{eq:drift},
because of inversion symmetry (see \ref{sub:Symmetries}). On the
other hand, Eq. \ref{eq:LangevinEquation} is equivalent to the forward
Fokker-Planck equation (in the present context, in which inertia forces are
neglected, it is also called Smoluchowski equation) for the probability
density $\rho\left(\mathbf{r},t\right)$ of finding the particle at
time $t$ in position $\text{\ensuremath{\mathbf{r}}}$, written as
\begin{eqnarray}
\frac{\partial\rho\left(\mathbf{r},t\right)}{\partial t} & = & -\Lambda_{\epsilon}\rho\left(\mathbf{r},t\right)\nonumber \\
\Lambda_{\epsilon}\rho\left(\mathbf{r},t\right) & = & -\frac{\epsilon}{2}\Delta\rho\left(\mathbf{r},t\right)+\mathbf{\nabla.}\left(\mathbf{f}\left(\mathbf{r}\right)\rho\left(\mathbf{r},t\right)\right),\nonumber \\
\rho\left(\mathbf{r},0\right) & = & \delta\left(\mathbf{r}-\mathbf{r}\left(0\right)\right).\label{eq:Fokker-Planck}
\end{eqnarray}

The evolution (Fokker-Planck) operator, $\Lambda_{\epsilon}$,
contains drift and diffusion terms: both are proportional to $\epsilon$
(the inverse of Peclet number). In the low noise limit, $\epsilon\rightarrow0$,
which is analyzed in the present work, a Brownian particle
stays most of its time exponentially close to the attractive stationary
points (SN). The noise term $\mathbf{\xi}\left(t\right)$ in Eq. \ref{eq:LangevinEquation}
allows for the particle to hop from one SN to another, if these \emph{metastable}
states are nearby ones. In fact, the low noise limit requires that,
asymptotically, these jumps should be to their first neighbor metastable
states. We concentrate from now on into cases A) through D) (see Table.
\ref{tab:Flow-types}). In each elementary cell $\left(-\pi,\pi\right)^{2}$,
we will identify the \emph{dominant} hopping process, in the sense
that it asymptotically dominates the contribution to the long-time
diffusion coefficient arising from Eq. \ref{eq:Diffusion}. This dominant
hop will be ``diagonal'' across the attractors located at $\mathbf{r}_{s1}=\left(2n+1,2m+1\right)\pi/2,\, n+m\in\mathrm{even}$.
The distance between these two adjacent (diagonal) attractors is $(2)^{1/2}\pi$
(in units of the inverse wavenumber $1/k$), so if the rate for this
process is called $r_{D}$, the diffusion coefficient can be
simply computed as
\begin{equation}\label{eq:DiffusionCoefficient}
D=2\pi^{2}r_{D}.
\end{equation}
In this low noise limit, all the hoppings between stable nodes (or
focus in case they existed) can be computed from the Mean First Passage
Time (MFPT) to the separatrix, which is the force line dividing the
domains of attraction of these SN. These separatrices are shown in
Fig. \ref{fig:grDriftCodesST}. The MFPT, $T\left(\mathbf{r}\right)$,
is defined as the average time spent by trajectories starting from position $\mathbf{r}$
before hitting any point in its closest separatrix for the first time. Asymptotically
$T\left(\mathbf{r}\right)\simeq T\left(\mathbf{r}_{s1}\right)$ for
$\epsilon\rightarrow0$ and all the hopping rates are given by $r_{D}\simeq1/[2T\left(\mathbf{r}_{s1}\right)]$
(see \citet{Talkner1987}). Hence, in this low noise limit and in
order to compute the effective diffusion constant in Eq. \ref{eq:Diffusion},
it is enough to compute the possible MFPT and chose the appropriate
one. How to obtain these paths, the corresponding passage times and which is the appropriate
one is the subject of next section.

\subsection{Low noise expansion.\label{sub:Low-noise-expansion.}}

From now on (unless otherwise stated) we will deal exclusively with
asymptotics for low noise, $\epsilon\rightarrow0$. For proofs and
motivation to the general arguments presented here, see \citet{Maier1997},
of which the present section is a quick summary of the needed results.

The MFPT, $T\left(\mathbf{r}_{s1}\right)$, to the separatrix $\partial\Omega$
of the domain of attraction $\mathbf{r}_{s1}\in\Omega$ is asymptotically
given by the smallest eigenvalue (all of them are positive) of the
Fokker-Planck operator (see Eq. \ref{eq:Fokker-Planck}) $\Lambda_{\epsilon}v_{\epsilon}^{0}=-\lambda_{\epsilon}^{\left(0\right)}v_{\epsilon}^{0}$,
with absorbing (Dirichlet) boundary conditions (i.e., indicating an
scape event ${\bf r}\in\delta\Omega$),
\begin{equation}
v_{\epsilon}^{0}\left(\mathbf{r}\in\partial\Omega\right)=0\;\mathrm{as}\; T\left(\mathbf{r}_{s1}\right)\simeq1/\lambda_{\epsilon}^{\left(0\right)}\label{eq:adsorbing}
\end{equation}
If $\mathbf{J}_{0}\left(\mathbf{r}\right)=\left[-\left(\epsilon/2\right)\mathbf{\mathbf{\nabla}+\mathbf{f}}\left(\mathbf{r}\right)\right]v_{\epsilon}^{0}\left(\mathbf{r}\right)$
is the probability current of this eigenstate, the eigenvalue $\lambda_{\epsilon}^{(0)}$
is the absorption rate, which just arises from the integral form of
the Fokker-Planck equation (Eq. \ref{eq:Fokker-Planck}) applied to its
eigenstate,
\begin{equation}
\lambda_{\epsilon}^{\left(0\right)}=\frac{\int_{\mathbf{r}\in\partial\Omega}\mathbf{J}_{0}\left(\mathbf{r}\right).\mathbf{n}\left(\mathbf{r}\right)}{\int_{\mathbf{r}\in\Omega}v_{\epsilon}^{0}\left(\mathbf{r}\right)},\label{eq:smallestEigenvalue}
\end{equation}
where $\mathbf{n}\left(\mathbf{r}\right)$ is the outward normal to
$\mathbf{r}\in\partial\Omega$. The distribution of exit points in
$\mathbf{r}\in\partial\Omega$ is $p\left(\mathbf{r}\right)\propto\mathbf{n}\left(\mathbf{r}\right).\mathbf{J}_{0}\left(\mathbf{r}\right)=-\left(\epsilon/2\right)\mathbf{n}\left(\mathbf{r}\right).\mathbf{\nabla}v_{\epsilon}^{0}\left(\mathbf{r}\right)$.

In solving $\Lambda_{\epsilon}v_{\epsilon}^{0}=\lambda_{\epsilon}^{\left(0\right)}$
and because of the exponential smallness of $\lambda_{\epsilon}^{\left(0\right)}$,
it is sufficient to consider the simpler $\Lambda_{\epsilon}v_{\epsilon}^{0}=0$
problem, and then, from Eq. \ref{eq:smallestEigenvalue} compute the
eigenvalue itself. The method to be employed is matched asymptotic
expansions, using an outer approximation, valid in most of $\mathbf{r}\in\Omega$,
based on a WKB type ansatz for the eigenstate,
\begin{equation}
v_{\epsilon}^{0}\left(\mathbf{r}\right)\simeq K\left(\mathbf{r}\right)\exp\left(-W\left(\mathbf{r}\right)/\epsilon\right).\label{eq:wkb}
\end{equation}
This outer solution does not obeys the adsorbing boundary condition,
but based on the form of $W({\bf r})$, we will show that for the
present drift, the only possible paths of escape from the SN's is
through saddles (SD). To complete the calculation of $v_{\epsilon}^{0}({\bf r})$
near the escape region, the outer solution has then to be matched
with a inner solution to the same equation ($\Lambda_{\epsilon}v_{\epsilon}^{0}=0$)
close to the saddles (SD). For this inner solution, which does obey
the boundary condition, it is enough to use the solvable linearized
drift approximation.

Construction of the outer approximation is involved when detailed
balance is lost. In the present constant diffusion Smoluchowski
case, it is equivalent to having forces not derivable from a potential
(this is the case we analyze in the present work). After substituting
the ansatz $v_{\epsilon}^{0}\left(\mathbf{r}\right)\simeq K\left(\mathbf{r}\right)\exp\left(-W\left(\mathbf{r}\right)/\epsilon\right)$
into $\Lambda_{\epsilon}v_{\epsilon}^{0}=0$ and making an expansion
in powers of $\epsilon$, the lowest order $O(1/\epsilon)$ leads to
an eikonal in the form of a zero energy Hamilton-Jacobi equation for
a generalized ``momentum'', given as the gradient of a function, ${\bf p}=\mathbf{\nabla}W\left(\mathbf{r}\right)$.
The Hamiltonian describes a generalized dynamics in a 4D phase-space,
with two spatial and two momentum coordinates,
\begin{eqnarray}
H\left(\text{\ensuremath{\mathbf{r}}},\mathbf{\nabla}W\left(\mathbf{r}\right)\right) & = & 0\nonumber \\
H\left(\text{\ensuremath{\mathbf{r}}},\mathbf{p}\right) & := & \frac{\mathbf{p}^{2}}{2}+\mathbf{p}.\mathbf{f}\left(\mathbf{r}\right)\label{eq:HamiltonJacobi}
\end{eqnarray}
This equation is solved by the method of characteristics which are,
precisely, Hamilton's equation of motion
\begin{eqnarray}
\dot{\mathbf{r}} & = & \mathbf{p}+\mathbf{f}\left(\mathbf{r}\right)\nonumber \\
\dot{\mathbf{p}} & = & -\nabla\left(\mathbf{p}.\mathbf{f}\left(\mathbf{r}\right)\right)\nonumber \\
\dot{W} & = & \frac{\mathbf{p}^{2}}{2},\label{eq:Hamilton}
\end{eqnarray}
In the last equation we have used that
\begin{equation}
\dot{W}=\dot{\mathbf{r}}. \nabla_{\mathbf{r}}W=\mathbf{p}\left(\mathbf{p}+\mathbf{f}\right)
\end{equation}
and that $H=0$ over the eikonal Eq. \ref{eq:HamiltonJacobi}, valid along
the characteristics given by the previous Hamilton equations. These
equations contain fluctuation free trajectories where $\mathbf{\nabla}W\left(\mathbf{r}\right)=\mathbf{p=0}$
and $\dot{\mathbf{r}}=\mathbf{f}\left(\mathbf{r}\right)$, which is 
just the deterministic drift equation. General arguments (see \citet{Maier1997})
show that Eqs. \ref{eq:Hamilton} have to be solved starting at $t=-\infty$
from the SN $\mathbf{r}_{s1}$, where we can fix $W\left(\mathbf{r}_{s1}\right)=0$, with no loss of generality.

It is important to note that the function $W\left(\mathbf{r}\right):=\int_{\mathbf{r}_{SN}}^{\mathbf{r}}\mathbf{p}.d\mathbf{r}$
is positive and locally single valued along any 2D manifold formed
with trajectories from this Hamiltonian flow (these manifolds are
termed Lagrangian, we will come back to this issue later). This is
key to determine the most probable escape path. This follows from the form
of the WBK ansatz for the probability $v_{\epsilon}^{(0)}$ in Eq.
\ref{eq:wkb}, which shows that the distribution of exit points is exponentially
dominated by the minimum of this positive function, $W\left(\mathbf{r}\right)$,
over the separatrix, $\mathbf{r}\in\partial\Omega$. The path which
joins $\mathbf{r}_{s1}$ to this minimum is called most probable escape
path (MPEP). In order to compute the MFPT from Eq. \ref{eq:smallestEigenvalue}
it is enough to consider $W\left(\mathbf{r}\right)$ close to this
escape point, and then match this WKB ansatz in a region of size $O\left(\epsilon^{1/2}\right)$
around this escape point to the inner approximation.

We will show that the local minima of $W({\bf r})$ we are interested
in are located at the saddle points (SD). From Eq. \ref{eq:HamiltonJacobi}
it is immediate that along the deterministic drift trajectories $\dot{\mathbf{r}}=\mathbf{f}\left(\mathbf{r}\right)$,
$\dot{W}\left(\mathbf{r}\right)=-|\mathbf{\nabla}W\left(\mathbf{r}\right)|^{2}/2\le0$,
thus showing that it is a Liapunov function for the drift (see \citet{Talkner1987}).
We note here that there is no contradiction between this last result
and $\dot{W}=\mathbf{p}^{2}/2\geq0,$ (from\ref{eq:Hamilton}), where
$W\left(\mathbf{r}\right)$ seems to only increase along trajectories.
One has to notice that the $W_{det}\left(\mathbf{r}\right)$ computed
with deterministic flow trajectories alone ($\mathbf{p=0}$) is null,
$W_{det}\left(\mathbf{r}\right)=0$, and that $\dot{W}=\mathbf{p}^{2}/2,$
is only valid along its defining trajectories.

\begin{figure}
\includegraphics[width=1\columnwidth]{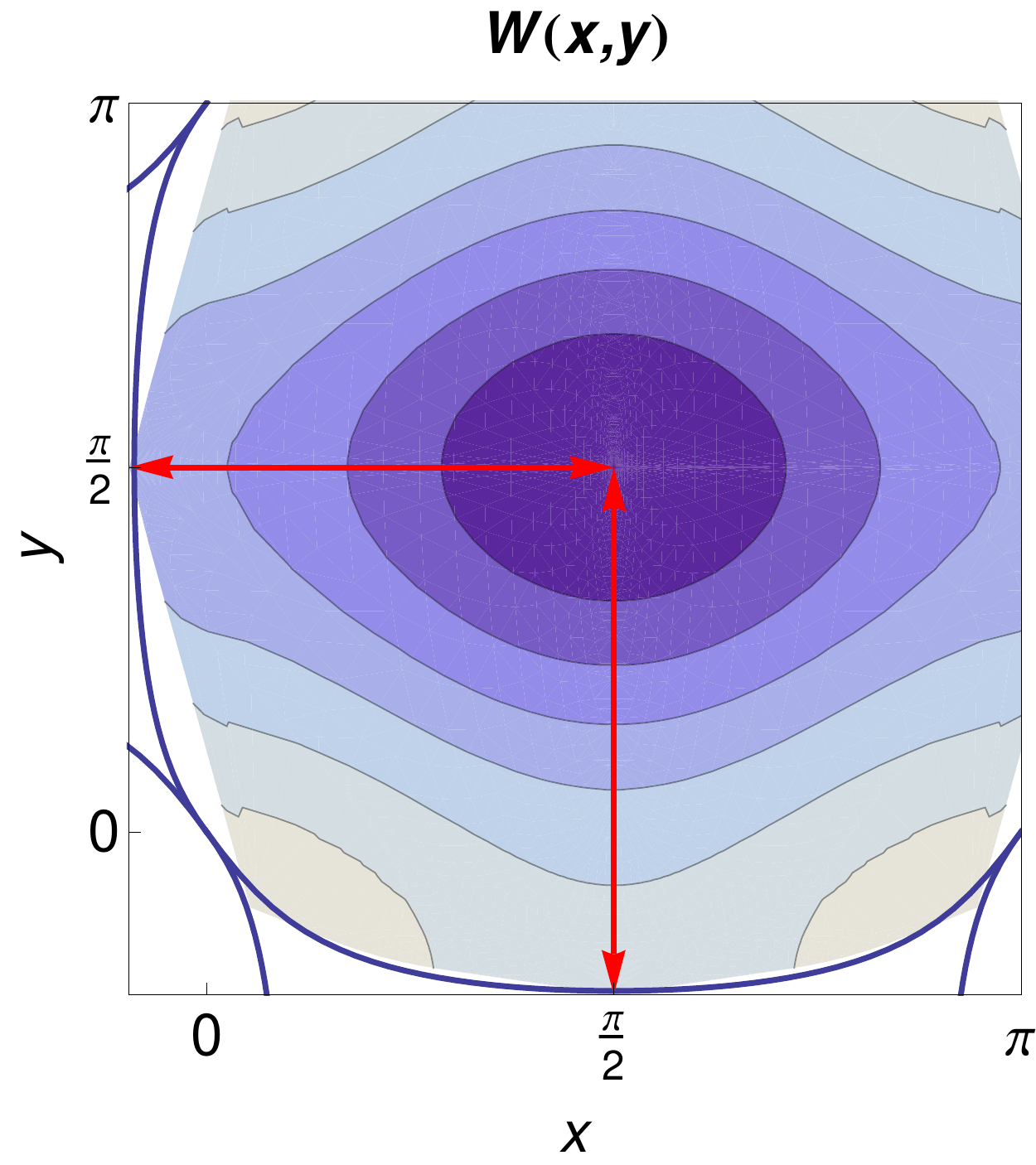}

\caption{\label{fig:grMPEPDensityPlot}Contours of $W\left(x,y\right)$ in an cell slightly larger than $\left[0,\pi\right]^{2}$ to accommodate the separatrices (in blue lines).
The parameters are taken from type A) in Fig. \ref{fig:grDriftCodesST}.
This shows (and will be justified below) that the MPEP (horizontal red arrowed line) is along the lines $y=\pi/2$, where $W\left(x,y\right)$ is appreciably smaller
than along the other MPEP (vertical red arrowed line) candidate lines $x=\pi/2$.}

\end{figure}

As the separatrices are deterministic trajectories, in which the drift
flows from the unstable (UF or UN) points at $\mathbf{r}_{s2}$ to
the saddles (SD) at $\mathbf{r}_{s3}$ , $\mathbf{r}_{s4}$ or $\mathbf{r}_{s1}$,
the former should be maxima an the later the minima. Hence, the only
candidates for the escape points are saddles (SD). There is a possibility
that actually $W\left(\mathbf{r}\right)$ is constant along the separatrix
(see \citet{Maier1992}) %
\footnote{This possibility requires a fine tuned field force such that every
trajectory (in the 3D zero energy surface in phase space), coming
from the stable node belonging to the unstable Lagrange manifold,
should arrive at all points in the separatrix with precisely zero
momentum, except for the saddles, which are not touched by any of
these trajectories.%
} however, in the present problem, this possibility can be ruled out
because we will explicitly build the trajectories joining the SN to
all the SD and because we will show that small fluctuations transverse
to the MPEP necessarily increase $W\left(\mathbf{r}\right)$. This
issue will be addressed later when we discuss the lack of foci along
the MPEP. To illustrate the discussion above, Fig. \ref{fig:grMPEPDensityPlot}
shows the form of $W({\bf r})$ in an elementary cell,
whose mimimum along the separatrix determines the most
probable escape path.

From Eq. \ref{eq:Hamilton} it is immediate to notice that the stationary
points of the drift, $\mathbf{r}_{s}$ (i.e ${\bf f}({\bf r}_{s})=0$),
lead to stationary points in the 4D phase space of the effective Hamiltonian
dynamics, $\left(\mathbf{r},\mathbf{p}\right)=\left(\mathbf{r}_{s},\mathbf{0}\right)$.
The linearized Hamilton equations near these fixed points read
\begin{eqnarray}
\frac{d}{dt}\left[\begin{array}{c}
\delta\mathbf{x}\\
\delta\mathbf{p}
\end{array}\right] & = & \left[\begin{array}{cc}
\mathbf{B}\left(\mathbf{r}_{s}\right) & \mathbf{I}\\
\mathbf{0} & -\mathbf{B}\left(\mathbf{r}_{s}\right)^{t}
\end{array}\right]\left[\begin{array}{c}
\delta\mathbf{x}\\
\delta\mathbf{p}
\end{array}\right]\nonumber \\
\mathbf{B}\left(\mathbf{r}_{s}\right)_{ij} & := & \frac{\partial\mathbf{f}\left(\mathbf{r}_{s}\right)_{i}}{\partial r_{j}},\label{eq:linearizedHamilton}
\end{eqnarray}
Diagonalization of the dynamic matrix of the linearized Hamiltonian
shows that, for each eigenvalue of the drift $\lambda_{i}$, the generalized
flow has $\pm\lambda_{i}$ as eigenvalues. This means that the 2D
manifold of zero energy trajectories starting at $t=-\infty$ from
the SN $\mathbf{r}_{s1}$ (where both eigenvalues of the drift are
negative) is the unstable 2D manifold of the generalized Hamiltonian
flow. On the other hand, the demand that the MPEP goes through saddles,
(which are also fixed points of the deterministic drift), implies first,
that this saddle is reached at infinite time, $t=\infty$ (precisely
because ${\bf p}=0$ at SD so the generalized force in Eq. \ref{eq:Hamilton}
is zero) and second, that this MPEP should also go through the stable
2D manifold through $\left(\mathbf{r}_{SD},\mathbf{0}\right)$. Hence,
the MPEP is the path which intersects these 2D manifolds. In the saddles,
if the stable/unstable eigenvalues of the drift are $\lambda_{s/u}^{SD}$,
then the Hamiltonian flow at $\left(\mathbf{r}_{SD},\mathbf{0}\right)$
possesses $\lambda_{s}^{SD}$ and $-\lambda_{u}^{SD}$ as stable eigenvalues,
with directions $\left(\mathbf{e}_{s},\mathbf{0}\right)$ and $\left(\tilde{\mathbf{e}}_{s},\tilde{\mathbf{g}}_{s}\right)$
respectively, and $\lambda_{u}^{SD}$ and $-\lambda_{s}^{SD}$ as
unstable eigenvalues, with directions $\left(\mathbf{e}_{u},\mathbf{0}\right)$
and $\left(\tilde{\mathbf{e}}_{u},\tilde{\mathbf{g}}_{u}\right)$
respectively (see Table \ref{tab:Unstable/stable--eigenvalues}).
The ratio of the size of these eigenvalues, $\mu:=|\lambda_{s}^{SD}|/\lambda_{u}^{SD}$
determines whether generically the MPEP goes through $\left(\mathbf{e}_{s},\mathbf{0}\right)$
for $\mu<1$ or though $\left(\tilde{\mathbf{e}}_{s},\tilde{\mathbf{g}}_{s}\right)$
for $\mu>1$ . Moreover, this ratio leads to different asymptotic
formula for the MFPT and qualitatively different exit point probability
distributions (see \citet{Maier1997} for these two issues). From
Tables \ref{tab:Unstable/stable--eigenvalues},\ref{tab:Flow-types}
it follows that $\mu>1$ always except for saddles of the type $\mathbf{r}_{s4}$
in cases A) and C), in which $\mu<1$. However, the symmetries of
the drift addressed in the present work lead to unmistakable identification
of all the MPEP candidates, and we find a situation in which these
MPEP go non-generically always through the directions $\left(\tilde{\mathbf{e}}_{s},\tilde{\mathbf{g}}_{s}\right)$.

The next order $O(\epsilon^{0})$ in the expansion in $\epsilon\ll1$
of the WKB approximation leads to the transport equation for the prefactor
$K\left(\mathbf{r}\right)$ along the characteristics $\mathbf{r}\left(t\right)$(the
initial condition can be chosen as $K\left(\mathbf{r}_{SN}\right)=1$)
\begin{equation}
\frac{dK\left(\mathbf{r}\left(t\right)\right)}{dt}=-\left[\mathbf{\nabla}.\mathbf{f}\left(\mathbf{r}\left(t\right)\right)+\frac{1}{2}tr\left(\frac{\partial^{2}W\left(\mathbf{r}\left(t\right)\right)}{\partial r_{i}\partial r_{j}}\right)\right]K\left(\mathbf{r}\left(t\right)\right),\label{eq:Transport}
\end{equation}
and a Riccati equation for the Hessian matrix $W_{,ij}:=\partial^{2}W\left(\mathbf{r}\left(t\right)\right)/\partial r_{i}\partial r_{j}$
(again along the characteristics $\left(\mathbf{r}\left(t\right),\mathbf{p}\left(t\right)\right)$):
\begin{equation}
\dot{W}_{,ij}=-W_{,ik}W_{,ik}-f_{k,j}W_{,ik}-f_{k,i}W_{,jk}-p_{k}f_{k,ij}\label{eq:TDRiccati}
\end{equation}
in which we have used Einstein summation convention for repeated indices
and, for example, $f_{k,i}:=\partial f_{k}\left(\mathbf{r}\left(t\right)\right)/\partial r_{i}$.
Near the attractors and the saddles, $\left(\mathbf{r}_{s},\mathbf{0}\right)$,
the relevant solution to this Ricatti equation has to be asymptotically
$t\rightarrow\pm\infty$ constant and the solution has to have rank
2. This, together with $\mathbf{B}\left(\mathbf{r}_{s}\right)$ being
diagonal and $\mathbf{p}=\mathbf{0}$ implies that
\begin{eqnarray}
\mathbf{Z}\left(\mathbf{r}_{s}\right)_{ij}: & = & \frac{\partial^{2}W\left(\mathbf{r}_{s}\right)}{\partial r_{i}\partial r_{j}}=-2\mathbf{B}\left(\mathbf{r}_{s}\right)_{ij}\label{eq:RiccatiSolutionNearStationaryPoints}
\end{eqnarray}

After matching to the inner approximation \cite{Maier1997} and, because
in Cartesian coordinates, all the matrices $\mathbf{B}\left(\mathbf{r}_{SD}\right)$
are diagonal and the drift-less diffusion coefficient matrix is diagonal
(here it is trivially $\epsilon\mathbf{I}$), the standard Eyring
formula (see \citet{Maier1997}) is valid for escape through then
saddles at $\mathbf{r}_{SD}$ (as a reminder, check that it is assumed that $W\left(\mathbf{r}_{SN}\right)=0$
and $K\left(\mathbf{r}_{SN}\right)=1$)
\begin{eqnarray}
& & T\left(\mathbf{r}_{SN}\right)^{-1} \simeq \lambda_{\epsilon}^{\left(0\right)} \nonumber \\
& \simeq & \frac{\lambda_{u}^{SD}}{\pi}\sqrt{\frac{|\det\mathbf{Z}\left(\mathbf{r}_{SN}\right)|}{\det\mathbf{Z}\left(\mathbf{r}_{SD}\right)}}K\left(\mathbf{r}_{SD}\right)e^{-W\left(\mathbf{r}_{SD}\right)/\epsilon}\nonumber \\
 & \simeq & \frac{1}{\pi}\sqrt{\frac{\lambda_{u}^{SD}\lambda_{1}^{SN}\lambda^{SN}{}_{2}}{|\lambda_{s}^{SD}|}}K\left(\mathbf{r}_{SD}\right)e^{-W\left(\mathbf{r}_{SD}\right)/\epsilon}\label{eq:Eyring}
\end{eqnarray}
here $\lambda_{s}^{SD},\,\lambda_{u}^{SD}$ are the stable, unstable
eigenvalues of the deterministic drift at $\mathbf{r}_{SD}$ and $\lambda_{1}^{SN},\,\lambda_{2}^{SN}$
the (both stable) eigenvalues at $\mathbf{r}_{SN}$. If there are
several non-equivalent saddles through which MPEP candidates can flow,
the ones with the smallest $W\left(\mathbf{r}_{SD}\right)$ are to
be chosen.

Implicit in the previous exposition is the assumption that along the
characteristics, $\left(\mathbf{r}\left(t\right),\mathbf{p}\left(t\right)\right)$,
in some region around the MPEP, the function $W\left(\mathbf{r}\right):=\int_{\mathbf{r}_{SN}}^{\mathbf{r}}\mathbf{p}.d\mathbf{r}$
is single valued, and therefore $W\left(\mathbf{r}_{SD}\right)$,
as computed from these MPEP trajectories, are local minima. This issue
has been thoroughly investigated in \citet{Maier1993,Maier1996} where
a criterion is given: the standard Jacobi criterion
which warranties that Hamilton's action, derived from the Lagrange
equations equivalent to Eq. \ref{eq:Hamilton}, is actually a minimum.
This Hamilton action is $W\left(\mathbf{r}\right)=\int_{-\infty}^{t_{1}}\left[|\mathbf{\dot{r}}\left(t\right)-\mathbf{f}\left(\mathbf{r}\left(t\right)\right)|^{2}/2\right]dt$
(also known by Wentzell-Freidlin action, and the Lagrangian, $L\left(\mathbf{r}\left(t\right),\mathbf{\dot{r}}\left(t\right)\right):=|\mathbf{\dot{r}}\left(t\right)-\mathbf{f}\left(\mathbf{r}\left(t\right)\right)|^{2}/2$,
as Onsager-Machlup). A Jacobi equation can be derived and numerically
solved to check whether Jacobi criterion is fulfilled. The answer
will turn to be positive, but we will address this issue in the next
section, where formula for the MPEP candidates will be available.

\subsection{Analytical results for the physical system. Lack of focusing.}\label{sec:analytical-results-for-the-physical-system.-lack-of-focusing.}

A glimpse at Fig. \ref{fig:grDriftCodesST} shows that all of the
possible MPEP candidates can be obtained by joining all the marked
SN, $\mathbf{r}_{SN}$, to their neighboring marked SD, $\mathbf{r}_{SD}$.
The rest lead to no new rates, by reasons of symmetry. Because
the deterministic flow $\mathbf{f}\left(\mathbf{r}\right)$ is vertical
along the vertical lines $x=\pi/2$, and horizontal along the horizontal
lines $y=\pi/2$, an analytical solution to Hamilton equations, which
obeys the boundary conditions $\mathbf{r}\left(t\rightarrow\infty\right)=\mathbf{r}_{SD}$
and $\mathbf{r}\left(t\rightarrow-\infty\right)=\mathbf{r}_{SN}$,
can be written straightforwardly
\begin{eqnarray*}
x & = & \pi/2\Rightarrow\begin{cases}
p_{x} & =0\\
\dot{y} & =-\left(\sin\left(2y\right)+2c_{+}\cos\left(y\right)\right)\\
p_{y} & =-2\, \left(\sin\left(2y\right)+2c_{+}\cos\left(y\right)\right)
\end{cases}\\
y & = & \pi/2\Rightarrow\begin{cases}
p_{y} & =0\\
\dot{x} & =-\left(\sin\left(2x\right)+2c_{-}\cos\left(x\right)\right)\\
p_{x} & =-2\, \left(\sin\left(2x\right)+2c_{-}\cos\left(x\right)\right)
\end{cases}
\end{eqnarray*}
The corresponding exponential factors $W\left(\mathbf{r}_{SD},\mathbf{r}_{SN}\right):=\int_{\mathbf{r}_{SN}}^{\mathbf{r}_{SD}}\mathbf{p}.d\mathbf{r}$
(we have changed a bit the notation to highlight the initial stationary
point), can be evaluated in closed analytical form, and results in:
\begin{align*}
 W&\left(\mathbf{r}_{s3},\mathbf{r}_{SN}=\left(\pi/2,-\pi/2\right)\right) =  2\left(1-c_{+}\right)^{2}\\
 W&\left(\mathbf{r}_{s4},\mathbf{r}_{SN}=\left(-\pi/2,\pi/2\right)\right) =  2\left(1-c_{-}\right)^{2}\\
 W&\left(\mathbf{r}_{s3},\mathbf{r}_{SN}=\left(\pi/2,\pi/2\right)\right) =  2\left(1+c_{+}\right)^{2}\\
 W&\left(\mathbf{r}_{s4},\mathbf{r}_{SN}=\left(\pi/2,\pi/2\right)\right) =  2\left(1+c_{-}\right)^{2}\\
 W&\left(\mathbf{r}_{SD}=\left(\pi/2,-\pi/2\right),\mathbf{r}_{SN} = \left(\pi/2,\pi/2\right)\right) = 8c_{+}.
\end{align*}

Using of Table \ref{tab:Flow-types} allows to sort the possible actions
for the different flow types. In cases A) and B), the MPEP for each
type of well is $W\left(\mathbf{r}_{s3},\mathbf{r}_{SN}=\left(\pi/2,-\pi/2\right)\right)<W\left(\mathbf{r}_{s4},\mathbf{r}_{SN}=\left(\pi/2,\pi/2\right)\right)$
and all the rest of escape paths are exponentially suppressed with
respect to these two. Because of this inequality, the particle will
spend most of the time in wells $\pm\left(\pi/2,\pi/2\right)$ and
diffusion will take preferentially from (say) $\left(\pi/2,\pi/2\right)\leftrightarrow-\left(\pi/2,\pi/2\right)$
via hopping to the intermediate $\pm\left(\pi/2,-\pi/2\right)$ wells,
where the particle stays a short time with respect to the long scale
given by the action $W\left(\mathbf{r}_{s4},\mathbf{r}_{SN}=\left(\pi/2,\pi/2\right)\right)=2\left(1+c_{-}\right)^{2}$.
Hence, at this long time scale, the diffusion process is diagonal.
Cases C) and D) lead also to the same MPEP, $W\left(\mathbf{r}_{s4},\mathbf{r}_{SN}=\left(\pi/2,\pi/2\right)\right)=2\left(1+c_{-}\right)^{2}$,
for the wells located at $\pm\left(\pi/2,\pi/2\right)$, but now,
because there is no stable points at the $\pm\left(\pi/2,-\pi/2\right)$
wells, the diffusion process proceeds directly $\left(\pi/2,\pi/2\right)\leftrightarrow-\left(\pi/2,\pi/2\right).$

Along the MPEP which dominates diffusion, namely, the path which joins
$\mathbf{r}_{SN}=\left(\pi/2,\pi/2\right)$ with $\mathbf{r}_{s4}=\left(-\arcsin c_{-},\pi/2\right)$,
the transport equation (Eq. \ref{eq:Transport}) can be immediately
solved as
\begin{eqnarray*}\label{eq:prefactor}
K\left(\mathbf{r}_{s4}\right) & = & \exp\left[\intop_{\pi/2}^{-\arcsin c_{-}}\frac{dx}{v_{0}\left(x\right)}\left(u_{1}\left(x\right)+\frac{w_{2}\left(x\right)}{2}\right)\right],
\end{eqnarray*}
where we have made a Taylor expansion around the MPEP (see \citet{Maier1996})
\begin{eqnarray*}
\mathbf{f}\left(x,y+\pi/2\right) & = & \left(\begin{array}{c}
v_{0}\left(x\right)+v_{2}\left(x\right)y^{2}+\ldots\\
u_{1}\left(x\right)y+\ldots
\end{array}\right)\Rightarrow\\
v_{0}\left(x\right) & = & 2\cos x\left(c_{-}+\sin x\right)\\
v_{2}\left(x\right) & = & -c_{-}\cos x\\
u_{1}\left(x\right) & = & -2\left(1+c_{+}\sin x\right)
\end{eqnarray*}
and $W\left(x,y-\pi/2\right)=w_{0}\left(x\right)+w_{2}\left(x\right)y^{2}/2+\ldots$
(here $W\left(\mathbf{r}\right):=W\left(\mathbf{r},\mathbf{r}_{SN}\right)$).
To get $w_{2}\left(x\right)$ we need to solve the Riccati equation
Eq. \ref{eq:TDRiccati} along the MPEP:
\begin{eqnarray}
v_{0}\left(x\right)w_{2}'\left(x\right) & = & w_{2}\left(x\right)^{2}+2u_{1}\left(x\right)w_{2}\left(x\right)-4v_{0}\left(x\right)v_{2}\left(x\right)\nonumber \\
w_{2}\left(\pi/2\right) & = & -2u_{1}\left(\pi/2\right)=2\left(1+c_{+}\right)\label{eq:Riccati1D}
\end{eqnarray}

\begin{figure*}
\includegraphics[width=2\columnwidth]{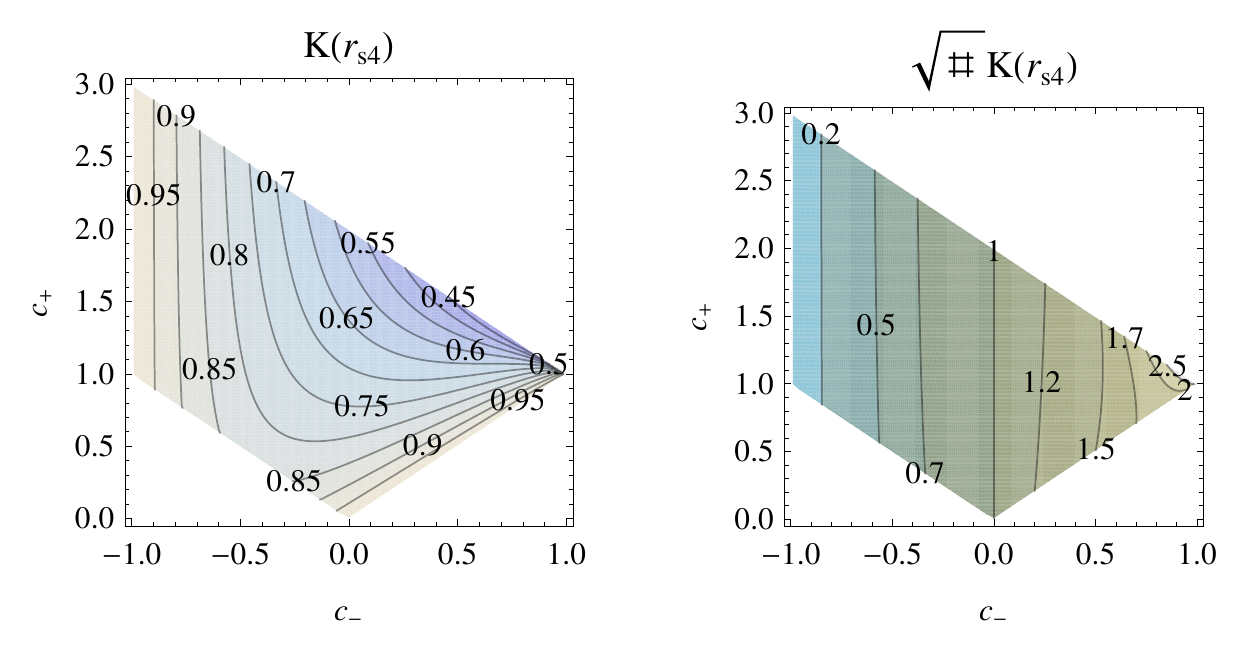}

\caption{\label{fig:grTransportPrefactorCountour}Contour lines for $K\left(\mathbf{r}_{s4}\right)$
and for the prefactor in Eq. \ref{eq:DiffusionResult}, where $\sqrt{\#}:=\sqrt{\left(1+c_{+}\right)\left(1+c_{-}\right)/\left(1-c_{+}c_{-}\right)}$.
This last figure shows a remarkable regularity, but we do not have
an argument in order to explain it.}
\end{figure*}

There is an important issue which could jeopardize the status of the
lines $y=\pi/2$ as MPEP and is related to them being a stationary
value of the action and not a minimum. The analysis in \citet{Maier1996}
reduces that problem to the analysis of a linear equation equivalent
to Eq. \ref{eq:Riccati1D}, which is a Jacobi equation, posed as
a boundary value problem. Whenever there is a solution, the pre-factor
$K\left(\mathbf{r}_{s4}\right)$ diverges, and the MPEP has bifurcated
into a symmetrically located pair of MPEP's (equivalently, a focus
appears at the end of the MPEP), which have lower value of $W\left(\mathbf{r}_{s4}\right)$
than that computed from the $y=\pi/2$ MPEP candidate. The solution
of the Jacobi equation, or the equivalent Riccati equation (Eq. \ref{eq:Riccati1D}),
is the only result whose solution cannot be obtained in closed analytical
form. However, it's numerical solution is straightforward. We found that
the $y=\pi/2$ lines minimize $W\left(\mathbf{r}_{s4}\right)$, i.e.,
the alluded bifurcation does not occur. Moreover, the dependence of
the pre-factor $K\left(\mathbf{r}_{s4}\right)$ on the external field
parameters $c_{-},c_{+}$ is very weak, as shown in Fig. \ref{fig:grTransportPrefactorCountour},
varying in the range $0.4\lesssim K\left(\mathbf{r}_{s4}\right)\le1$.

Collection all into the Eyring formula Eq. \ref{eq:Eyring}, the diffusion
coefficient is asymptotically equal to:
\begin{align}
D\simeq & \pi\sqrt{\frac{1-c_{-}^{2}}{1-c_{+}c_{-}}\left(1+c_{+}\right)\left(1+c_{-}\right)}K\left(\mathbf{r}_{s4}\right)\nonumber \\  &\exp\left[-\frac{2}{\epsilon}\left(1+c_{-}\right)^{2}\right]\label{eq:DiffusionResult}
\end{align}

Numerically, as seen in Fig. \ref{fig:grTransportPrefactorCountour}, we find that
the total temperature independent prefactor (all except the exponential)
in Eq. \ref{eq:DiffusionResult} varies in the approximate range $0.014-3.47$.

\subsection{Conservative systems\label{sub:Conservative-systems}}

The cases in which only gradient forces appear can be solved similarly,
but some simplifications happen, namely, the pre-factor is trivial,
$K\left(\mathbf{r}\right)=1$, and the exponent $W\left(\mathbf{r}\right)=2V\left(\mathbf{r}\right)$,
where $\mathbf{f}\left(\mathbf{r}\right)=-\nabla V\left(\mathbf{r}\right)$.
There are two possibilities, either $\beta=0,\,\phi\ne0$ or $\phi=0$.
The former is, from the point of view of the drift topology, a pure
degenerate case A) with the particularity that now vertical and horizontal
hopping rates are the same. Taking this into account, one immediately
finds twice the same expression as in Eq. \ref{eq:DiffusionResult},
due to the doubling of the hopping processes

\begin{equation}
D\simeq2\pi\left(1+\cos\phi\right)\exp\left[-\frac{2}{\epsilon}\left(1+\cos\phi\right)^{2}\right]\label{eq:DiffusionGradientA}
\end{equation}

The $\phi=\pi/2$ possibility should not be included in this formula
because now, all the $\mathbf{r}_{s1}=\left(2n+1,2m+1\right)\pi/2$
points are at the same potential level and, therefore, all the horizontal
and vertical transitions among neighbors of this type occur at the
same rate. On the other hand, this is a soluble case because the Langevin
equations are now separable and an analytical exact formula (see \citet{Risken1984})
is available in terms of the Bessel function $I_{0}\left(x\right)$:
\begin{equation}\label{eq:DiffusionGradientB}
D=\frac{\epsilon}{2}I_{0}\left(\frac{1}{\epsilon}\right)^{-2}\simeq\pi e^{-2/\epsilon},
\end{equation}
where in the last part we have exhibited the low noise ($\epsilon\rightarrow0$)
asymptotic.

The other $\phi=0$ case has to be worked out, because now the lines
$x+y=2n\pi$ and $x-y=2(n+1)\pi$ (with $n$ integer) are equi-potential.
In these cases, the prefactor is not independent of $\epsilon$ but
the computation is straightforward. Adapting the results of \citet{Matkowsky1983}
to this case one gets
\begin{equation}\label{eq:DiffusionGradientC}
D\simeq4\sqrt{\frac{2\pi}{\epsilon}}e^{-8/\epsilon}
\end{equation}

\subsection{Regimes without stable nodes}\label{sec:no-stable-nodes}
The diffusion coming from case E) dynamics (see Fig. \ref{fig:grDrifTypeE}) requires a different though more straightforward approach. It is not difficult to show that the lines $x,y=(2n+1)\pi/2$ are heteroclinic orbits which constitute the $\omega$-limit set of the flow. As the velocity field goes to zero when approaching the saddles at the vertices, one concludes that, when the amount of noise is small $\epsilon\ll 1$, particles will spend most of the time in a 2D region of area $\sim \epsilon$ near those vertices at $\mathbf{r}_{s1}$, where the flow can be linearized. We will show that the actual size of that region is not important if the leading low noise behavior is sought.

One is then confronted to find the MFPT from a region of area $\sim \epsilon$ to some 1D border located a distance $\sim 1$ from the points $\mathbf{r}_{s1}$. Again, it will be shown that the precise distance is of no concern.

There is a related approach to compute the MFPT, based on Kolmogorov's backward equation, (see \cite{Matkowsky1983}), which in this case amount to solving the boundary value problem for the MFPT, $T(\mathbf{r})$ from $\mathbf{r}$ to the border $\Omega$
\begin{eqnarray}
-1 & = & \frac{\epsilon}{2}\Delta T\left(\mathbf{r}\right)+\mathbf{f}\left(\mathbf{r}\right)\mathbf{.\nabla}T\left(\mathbf{r}\right), \nonumber \\
0 & = & T\left(\mathbf{r} \in  \Omega\right).\label{eq:KolmogorovBE}
\end{eqnarray}

\begin{figure*}
\includegraphics[width=\columnwidth,angle=90]{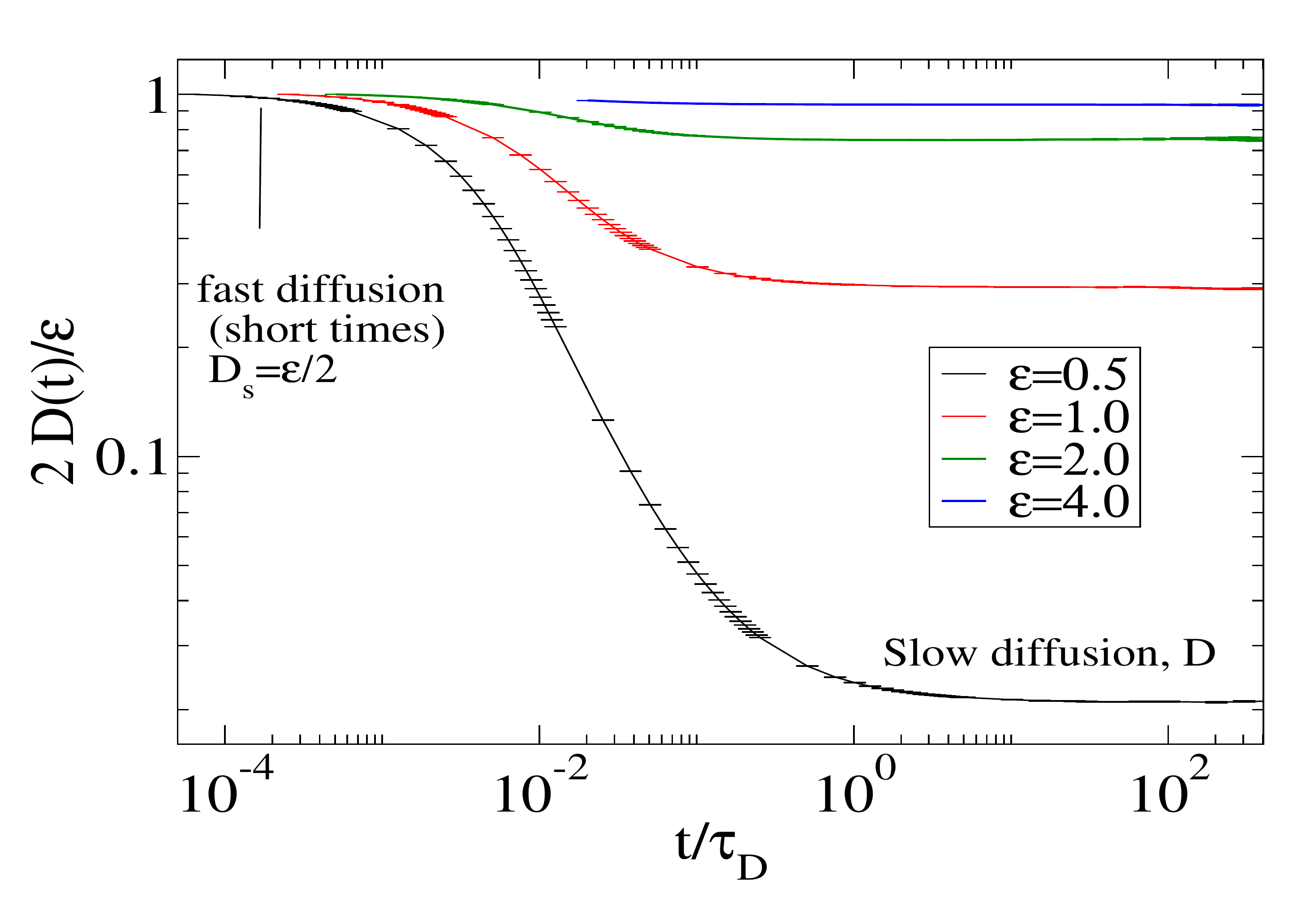}
\caption{\label{fig:grDiffusionRegimes} The time dependent diffusion coefficient
(particle mean square displacement scaled with time), showing the two
distinct dynamic regimes. At short times nanoparticles are wandering
around the stable node with a diffusion coefficient given by $\epsilon/2$,
at large times, the (slow) diffusion is controlled by the transition
rate for the particle to jump to a neighboring SN. Numerical results
correspond to parameter of Case A of Fig. \ref{fig:grDriftCodesST}.}
\end{figure*}

For a 1D Ornstein-Uhlenbeck Process (OUP), an analytical solution is available. We will need that, in this case, i.e. for a force $f(x)=-\alpha x$, the solution of Eq. \ref{eq:KolmogorovBE} for the symmetrical border situated at $x=\pm a$ is
\begin{eqnarray}
T(x) & = & \frac{1}{\epsilon} \left[a^2 H\left(\frac{a^2 \alpha}{\epsilon}\right) - x^2 H\left(\frac{x^2 \alpha}{\epsilon}\right) \right] \nonumber \\
H(z) &:= & _pF_q \left((1,1),(\tfrac{1}{2},2);z\right),
\end{eqnarray}
and $_pF_q $ is a Generalized Hypergeometric Function. For an attractive 1D-OUP, $\alpha>0$, the low noise asymptotic expansion of $T(\lambda a)\simeq \sqrt{\pi \epsilon /4a^2 \alpha^3} \exp(a^2 \alpha/\epsilon)$ is independent of $0\le\lambda<1$, which is exponentially large. On the other hand, for a repulsive 1D-OUP, $\alpha<0$, and for a region of size $\lambda=\sqrt{\epsilon\delta}$, where $\delta \sim 1$, we get $T(\lambda a)\simeq \log \epsilon /2 \alpha$, which is now independent of both $\delta, a$. This means, as it should be evident on physical grounds, that for the 2D-OUP which comes from the linearization of the flow around the vertices $\mathbf{r}_{s1}$, one can forget the processes which leave any squared region of area $\sim 1$ through the attractive border, and that the 2D-OUP MFPT shows this last asymptotic behavior, which is independent of both the size over which particles expend most of their time or the distance to the artificial boundary to define the MFPT. But now, one has also to consider how long does it take for the particles to reach a region of linear size $\sqrt{\epsilon \delta'}, \delta'\sim 1$ around the stationary points. This is a deterministic problem whose solution is the same as the one obtained for the MFPT, i.e. this time is $\simeq |\log(\epsilon)|/2\alpha'$, where now $\alpha'>0$ comes from the  attractive part of the force.

Adapting the discussion around Eq. \ref{eq:DiffusionCoefficient}, the low noise leading behavior of the diffusion coefficient is
\begin{eqnarray}
D & \simeq & \frac{\pi^2}{2|\log \epsilon|}\left[(\lambda_{u_+}^{-1}-\lambda_{s_+}^{-1})^{-1} + (\lambda_{u_-}^{-1}-\lambda_{s_-}^{-1})^{-1}\right] \nonumber \\
\lambda_{u_\pm} & := & 2(\pm c_\pm -1) \nonumber \\
\lambda_{s_\pm} & := & -2(1\pm c_\pm),
\end{eqnarray}
being $\lambda_{u_\pm},\lambda_{s_\pm}$ the unstable/stable eigenvalues of the linearized force field at the saddles, see Eq. \ref{eq:linearizedHamilton} and Table \ref{tab:Unstable/stable--eigenvalues}.

\subsection{Comparison with numerical Langevin simulations.}\label{sec:comparison-with-numerical-langevin-simulations.}

\begin{figure*}
\includegraphics[width=8cm,angle=90]{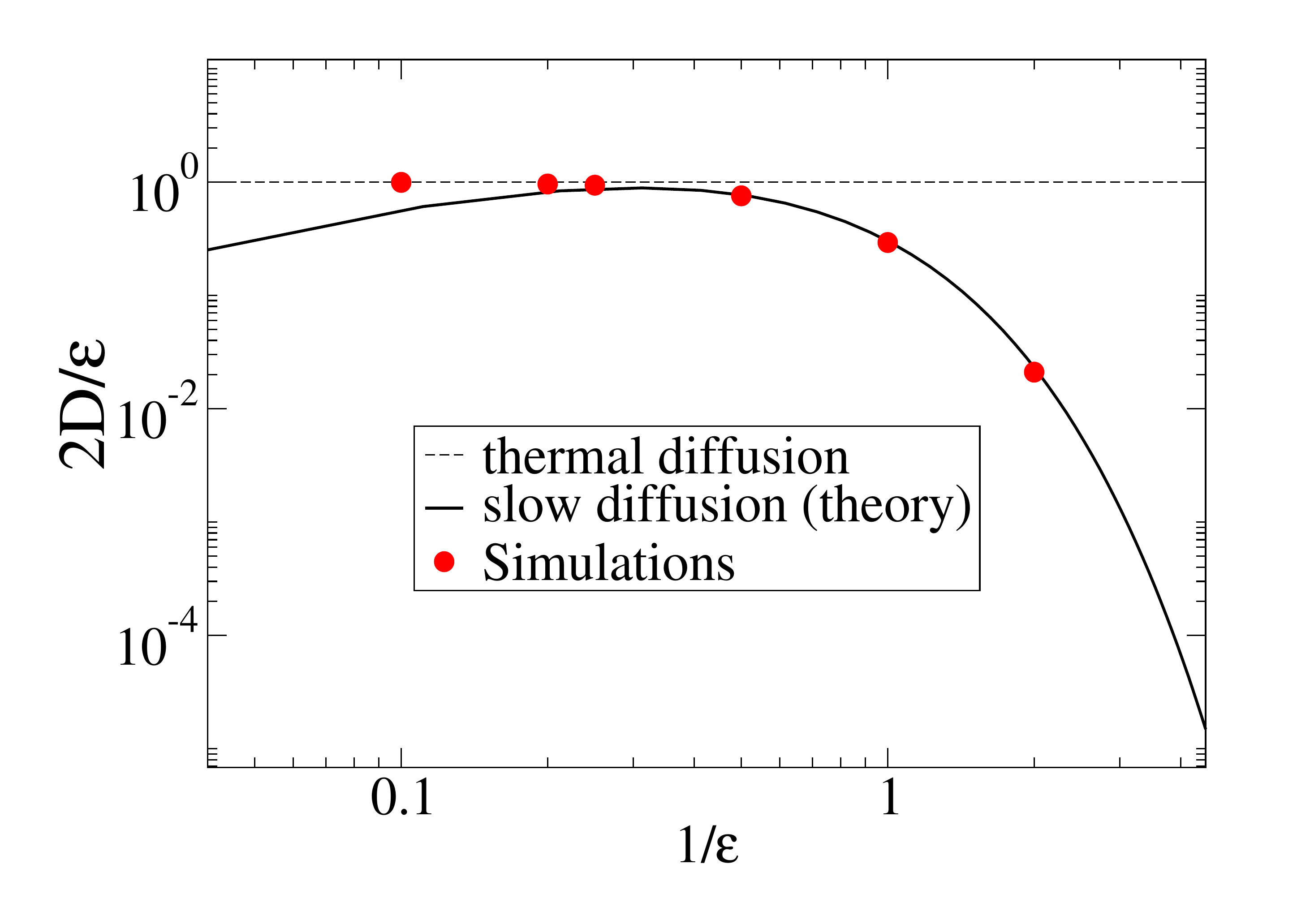}
\caption{\label{fig:grDiff_vs_eps} Long time diffusion coefficient corresponding
to Case A of Fig. \ref{fig:grDriftCodesST}. Circles are results from
numerical simulations of the Langevin equation. The solid line is
the analytical result of Eq. \ref{eq:DiffusionResult} and the
dashed line indicates the fast diffusion coefficient $\epsilon/2$.
At large $\epsilon$, thermal diffusion overpowers the effect of optic
forces.}
\end{figure*}

We have numerically solved the particle dynamics described by Eq.
\ref{eq:LangevinEquation} to assess the quality of the theoretical
predictions for the long time diffusion coefficient based on the low
noise approximation ($\epsilon\ll1)$. We illustrate here results
for one of the cases considered type A) of Table \ref{tab:Flow-types},
with parameters as in Fig \ref{fig:grDriftCodesST}, namely, $\beta=0.7,\,\phi=0.7\pi/2$.
To solve the over-damped Langevin dynamics of \ref{eq:LangevinEquation}
we have used an explicit (first order) Euler method with a small enough
time step $\Delta t=0.01$, which is just enough to deal with the
moderate computational weight of the present test. However, sampling
the long-time regime requires extremely long simulations as $\epsilon\rightarrow0$
and a complete study of the validity of the asymptotic analysis presented
above would require more accurate schemes, such as Heun or higher
order Runge-Kutta schemes for stochastic ordinary differential equations.
The mean square displacement of the particle, $\langle(\mathbf{r}(t)-\mathbf{r}(0))^{2}\rangle$
is shown in Fig.\ref{fig:grDiffusionRegimes} for several values of
the noise amplitude $\epsilon$ parameter. Two regimes are clearly
distinguished: at short times the diffusion is controlled by the fast
dynamics of Eq. \ref{eq:LangevinEquation}, whose diffusion coefficient
is $D_{s}=\epsilon/2$ (see the associated Fokker-Planck equation
\ref{eq:Fokker-Planck}). At long times the dynamics is governed by
the non-conservative drift term of the Fokker-Planck operator. Figure \ref{fig:grDiffusionRegimes}
shows that the transition from the short to the long time dynamical
regime is relatively fast, and it becomes sharper as the noise is
reduced (small values of $\epsilon).$ This reflects the fast process
of jumping from one SN to another and it is in agreement with the
theoretical description given above. The asymptotic value of the
diffusion coefficient in the long time dynamics $D=D(\epsilon)$ is
controlled by the hopping rate between stable nodes, determined by
both, the force landscape and the noise amplitude. In the Case A,
the long time diffusion coefficient is predicted by Eq. \ref{eq:DiffusionGradientA}
which is expected to be exact in the limit $\epsilon\rightarrow0$.
Figs. \ref{fig:grDiff_vs_eps}
compare the theoretical prediction with the numerical estimations
of $D(\epsilon)$. Remarkably, the asymptotic limit $\epsilon\rightarrow0$
provides an excellent prediction which holds true up to $\epsilon\lesssim0.4$.
At larger values of the noise parameter, thermal fluctuations become
too large for the particle to ``see'' any detail of the optical
force landscape.

\begin{figure*}
\includegraphics[width=2\columnwidth]{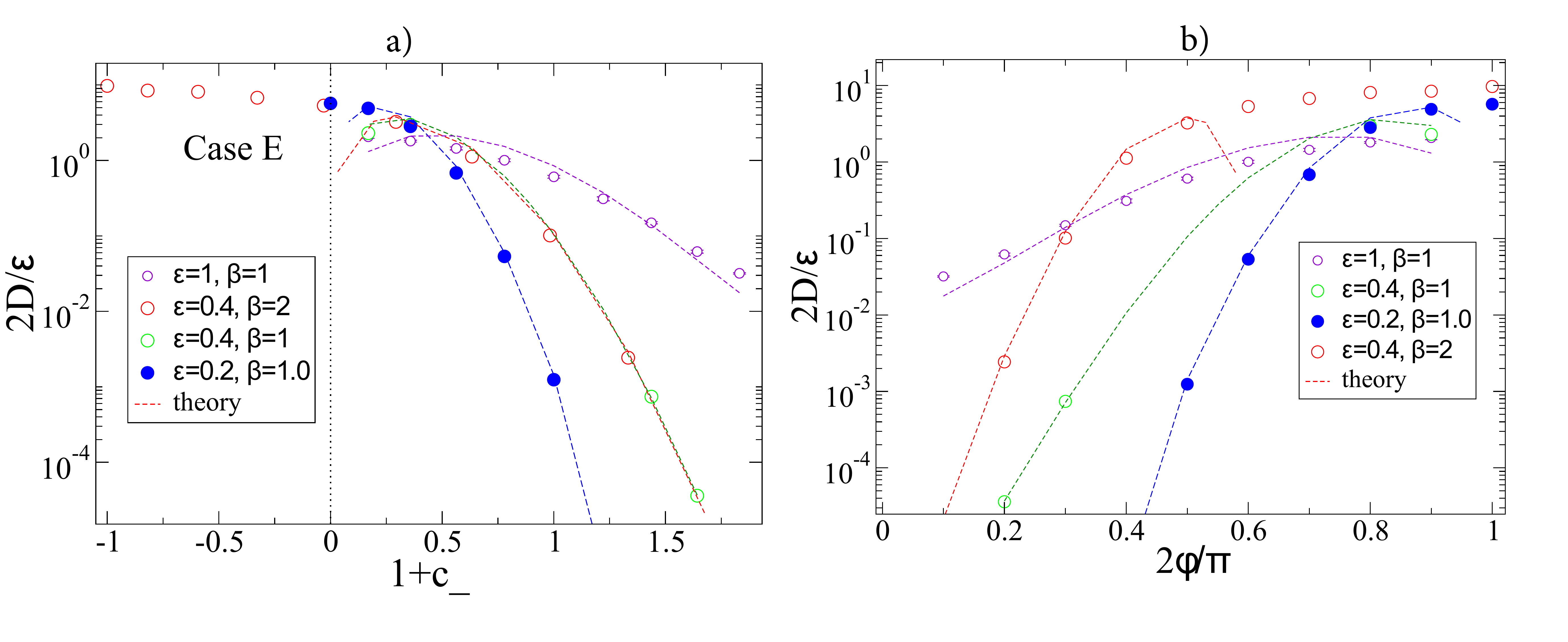}
\caption{\label{fig:grLongDiffScaled_eps} The long time diffusion scaled with the thermal diffusion coefficient
($2D/\epsilon)$ for a set of parameters of the setup. Dashed lines
corresponds to the theoretical result of Eq. \ref{eq:DiffusionResult}
and circles to simulations (same color code as lines). In a) we present
results against $(1+c_{-})$, showing an excellent agreement with
theory. It is noted that, for the regime treated in this work $(1+c_{-})>0$,
such scaling leads to $\beta$-independent trends. In b) we plot the
scaled diffusion against the lasers' phase shift $\phi$, illustrating
the huge jump in diffusivitites this setup allows to control.}
\end{figure*}

We performed another set of simulations to further investigate the
range of validity of the main result of this paper, Eq. \ref{eq:DiffusionResult}.
This is an Arhenius-type relation, originating from the WBK approximation
which usually leads to unexpectedly wide ranges of validity. In particular,
for Eq. \ref{eq:DiffusionResult} to be valid, the condition $2\left(1+c_{-}\right)^{2}/\epsilon\gg1$
should hold. Figure \ref{fig:grLongDiffScaled_eps} compares the prediction of Eq. \ref{eq:DiffusionResult}
and simulations for a range of values of $\epsilon,\,\beta$ and phase shift
angle$\phi$. The case $\beta=2$ is consistent with gold nanoparticles
and has been considered in Ref. \cite{Albaladejo2009}, where Case E ($\beta>1,\:\phi=\pi/2$)
was studied. In order to illustrate the global picture, we have extended
the numerical analysis to this Case E (see the vector-field of Fig.\ref{fig:grDrifTypeE}),
which corresponds to a dynamical regime where the long time diffusion
is substantially enhanced, and not reduced as in cases A to D (see
Fig. \ref{fig:grDriftCodesST}). Case E will be considered in future numerical
efforts. The first conclusion drawn from Fig. \ref{fig:grLongDiffScaled_eps}a is the relatively
large range of validity of the present theory. Comparing with numerical
results, we conclude that Eq. \ref{eq:DiffusionResult} is valid for
$2\left(1+c_{-}\right)^{2}/\epsilon>0.2$ and (provided this condition
holds) it even predicts the correct trend for $\epsilon=1$. A second
conclusion concerns the applicability of the present problem in the
control and manipulation of metal nanoparticles by optical forces.
In particular, Fig. \ref{fig:grLongDiffScaled_eps}b presents the nanoparticle long time diffusion
against the phase-lag angle $\phi$ between the two lasers. Interestingly,
for a fixed laser power, one can tune the phase lag to pass immediately
from a giant particle diffusion about 100 larger than the thermal
one (using$\phi=\pi/2$) to a frozen state, where particles are trapped
in deep wells and do not diffuse at all (at small enough $\phi$).
We believe that such control over these completely opposite dynamics,
with a huge jump in the nanoparticles diffusion coefficient, will
allow for new and interesting applications.

\section{Conclusions}\label{sec:conclusions}
In this work we have analyzed the diffusion of a 2D system of nano-particles subject to a definite optical forces field which leads to a non-conservative system. The results obtained for the diffusion coefficient in the low-noise regime is an almost complete analytical result, in the sense that all that requires numeric computation is a single function, $K\left(\mathbf{r}_{s4}\right)$ (see Eqs. \ref{eq:prefactor}, \ref{eq:DiffusionResult}) of the two force field parameters $c_{+}, c_{-}$, which can be easily computed by solving a non-linear first order Riccati equation, Eq. \ref{eq:Riccati1D} and tabulated (as shown in Fig. \ref{fig:grTransportPrefactorCountour}). As shown in Sec. \ref{sec:comparison-with-numerical-langevin-simulations.} the agreement with Langevin simulations is more than satisfactory.

One might wonder whether the present setup could be slightly adapted in order to show the other phenomena deduced in \cite{Maier1993,Maier1996,Maier1997}, namely, focusing, caustics and wedge regions classically forbidden (some discussion was given in the introduction and in Sec. \ref{sec:analytical-results-for-the-physical-system.-lack-of-focusing.}). Indeed, as follows from the discussion following Eq. \ref{eq:linearizedHamilton}, cases A) and C) have the peculiarity that $\mu<1$ at the saddles $\mathbf{r}_{s4}$, which are the endpoints of the MPEP lines $y=\pi/2$. But it is the strong symmetry of the present force field along these lines which makes them non-generically MPEP, as discussed in \cite{Maier1997}. Modifications of the force field can qualitatively change these MPEP near the saddles so that the wedge regions show their effect on the diffusion properties. The addition of other laser fields allows to engineering of the force field so that focusing and caustics may be forced to occur. All these is the subject of a future work. 


\begin{acknowledgments}
We thank Juan Manuel Rodríguez Parrondo, Manuel I. Marqués, Jorge Luis Hita and Peter
Reinmann for useful discussions. This work has been supported by Comunidad de Madrid through grant MICROSERES-CM (No. S2009-TIC-1476), the Spanish Ministry of Economy and Competitiveness (grant numbers FIS2012-36113-C03,  FIS2013-41716-P and Proyecto Explora NANOMIX FIS2013-50510-EXP).
\end{acknowledgments}

\bibliographystyle{apsrev4-1}
\bibliography{DiffusionControl_10}

\end{document}